\documentclass[11pt]{article}
\usepackage[margin=1in]{geometry}
\usepackage{graphicx}
\usepackage[numbers,sort]{natbib}
\usepackage{footnote,enumerate,amsmath,amssymb,amsfonts,amsthm,subcaption,hyperref}

\usepackage[linesnumbered,ruled]{algorithm2e}
\newcommand{\R}{{\mathbb R}}

\begin{document}
\title{Efficient Combinatorial Optimization Using Quantum Annealing}
\author{Hristo Djidjev (PI)\\Los Alamos National Laboratory
\and Guillaume Chapuis\\Los Alamos National Laboratory
\and Georg Hahn\\Imperial College, London, UK
\and Guillaume Rizk\\INRIA, Rennes, France}
\date{November 8, 2016}
\maketitle

\begin{abstract}
The recent availability of the first commercial quantum computers has provided a promising tool to tackle NP hard problems which can only be solved heuristically with present techniques. However, it is unclear if the current state of quantum computing already provides a quantum advantage over the current state of the art in classical computing. This article assesses the performance of the D-Wave 2X quantum annealer on two NP hard graph problems, in particular clique finding and graph partitioning. For this, we provide formulations as Qubo and Ising Hamiltonians suitable for the quantum annealer and compare a variety of quantum solvers (Sapi, QBSolv, QSage provided by D-Wave Sys, Inc.) to current classical algorithms (METIS, Simulated Annealing, third-party clique finding and graph splitting heuristics) on certain test sets of graphs. We demonstrate that for small graph instances, classical methods still outperform the quantum annealer in terms of computing time, even though the quality of the best solutions obtained is comparable. Nevertheless, due to the limited problem size which can be embedded on the D-Wave 2X chip, the aforementioned finding applies to most of problems of general nature solvable on the quantum annealer. For instances specifically designed to fit the D-Wave 2X architecture, we observe substantial speed-ups in computing time over classical approaches.
\end{abstract}

\section{Introduction}
\label{section_introduction}
The availability of the first commercial quantum computers \citep{dwave2016} has led to many new challenges in the areas of computing, physics and algorithmic design. In particular, the new devices provide a promising tool to solve NP hard problems which can only be solved heuristically with present techniques. However, it is still disputed in the literature if the available quantum computers indeed provide an advantage over classical methods \citep{Ronnow2014,Denchev2016}.

This article investigates the performance of the D-Wave 2X quantum computer on two NP hard graph problems, maximal clique finding and (edge-cut as well as \textit{ch}-) partitioning. The first problem aims to find the largest complete (i.e., fully connected) subset of vertices for a given graph \citep{Balas1986}. The latter typically aims to divide a graph into a pre-specified number of components in such a way that the number of edges in-between all components is minimized \citep{Buluc2015}. Another variant of graph partitioning investigated in this article is \textit{core-halo} (CH) partitioning by \cite{Djidjev2016}.

A variety of classical heuristics are known to solve both problems in reasonable time with high accuracy. The device developed by D-Wave Systems, Inc., is a quantum annealer which follows a different approach. Briefly speaking, a quantum annealer is a hardware realization of the classical simulated annealing algorithm, with the exception that it makes use of quantum effects to explore an energy or objective function landscape. Similarly to the random moves proposed in simulated annealing, quantum annealers use quantum fluctuations to escape local minima and to find low-energy configurations of a physical system. The quantum bits (qubits) are realized via a series of superconducting loops on the D-Wave chip. Each loop encodes both a 0 and 1 bit at the same time through two superimposed currents in both clockwise and counter-clockwise directions until annealing is complete and the system turns classical \citep{Johnson2011,Bunyk2014}.

Due to the particular architecture of the D-Wave chip, the device is designed to minimize an unconstrained objective function consisting of a sum of linear and quadratic qubit contributions, weighted by given constants. To be precise, it aims to minimize the Hamiltonian
$$\sum_{i \in V} a_i q_i + \sum_{(i,j) \in E} a_{ij} q_i q_j$$
for given $a_i$, $a_{ij} \in \mathbb{R}$, where $V = \{1,\ldots,N\}$ and $E=V \times V$ \citep{King2015}. The advantage of a quantum computer for minimizing an Hamiltonian of the aforementioned form lies in the fact that, although classically an NP hard combinatorial problem, its mapping of the problem onto a physical system allows to read off solutions by observing how the system arranged itself in an (energy) minimum. Notably, the computing time is independent of the problem size and unlike existing digital computers, which are basically deterministic, quantum annealing is fundamentally probabilistic: Due to the experimental nature of the minimisation procedure, solutions are not always identical.

The article is structured as follows. Section \ref{section_methods} starts by introducing the Qubo (quadratic unconstrained binary optimization) and Ising Hamiltonians and contains a note on the D-Wave architecture. Moreover, it describes the quantum tools (\textit{Sapi}, \textit{QBsolv} and \textit{QSage}, see \cite{DWaveDevelopperGuide2016}) as well as the classical algorithms (simulated annealing \citep{Kirkpatrick1983}, \texttt{METIS} \citep{KarypisKumar1999} and \texttt{Gurobi} \citep{gurobi}) we make use of in this study.

Section \ref{sec:maxc} investigates the maximum clique problem. It describes its Qubo formulation, a graph splitting algorithm as well as the parameters we used to solve the maximum clique problem on D-Wave. Moreover, Section \ref{sec:maxc} presents a comparison study of D-Wave to for small graphs as well as graphs specifically chosen to fit the D-Wave qubit architecture. The comparison involves classical methods (quantum adiabatic evolution of \cite{childs2000finding}, specialized simulated annealing of \cite{geng2007simple}), the \texttt{fmc} clique heuristic and exact solver \cite{pattabiraman2013fast} and \texttt{Gurobi}) as well as the aforementioned solvers provided by D-Wave.

Section \ref{section_graph_partititioning} focuses on two flavors of graph partitioning, called edge-cut and CH-partitioning. It derives Hamiltonians in both Qubo and Ising formulation and proves conditions on the Hamiltonian weights which guarantee that the global minimum of the Hamiltonian is attained at the optimal partitioning solution. Using random graphs, Section \ref{section_graph_partititioning} evaluates all the aforementioned classical and quantum solvers in terms of computing time as well as the quality of the partitionings computed, both on the edge-cut and on the CH-partitioning problem. We also compare the embedding time vs.\ the anneal time on the D-Wave device.

We conclude with a discussion of our results in Section \ref{section_conclusion}, arguing that a considerable quantum advantage is observable in instances only which are tailored to the D-Wave chip, but not in problem instances of general form.

In the entire paper, we denote a graph as $G=(V,E)$, consisting of a set of $n$ vertices $V = \{ 1,\ldots,n \}$ and a set of undirected edges $E$. The cardinality of a set $V$ is denoted as $|V|$.

\section{Methods}
\label{section_methods}
This section introduces quadratic integer programs, the type of problem the D-Wave device is capable of solving. We provide two different formulations of such quadratic programs, and show how to convert between the two. Next, this section briefly presents three tools provided by D-Wave Inc.\ to submit quadratic programs to the quantum computer. To this end, we treat their input format, the solver used and its interface, as well as available post-processing methods and output format. The section concludes with a summary of classical methods we will use in the forthcoming sections to compare any D-Wave results.

\subsection{Ising and Qubo formulations}
\label{section_ising_qubo}
The D-Wave device is designed to solve a certain type of problem only, called quadratic integer programs. It operates on $n$ variables $q_i$, $i \in \{1,\ldots,n\}$, and allows linear and quadratic interactions, summarized as
\begin{align}
\min_{q_1,\ldots,q_n} \Big(\sum_{i=1}^n a_i q_i + \sum_{1\leq i<j\leq n} \!\!\!a_{ij}\,q_i q_j\Big)
\label{eq:quadratic_program}
\end{align}
where the linear weights $a_i \in \R$ and the quadratic coupler weights $a_{ij} \in \R$ are pre-specified by the user. The aforementioned quadratic minimization program exists in two flavors: In a \textit{Qubo} formulation, in which all variables take binary values, i.e.\ $q_i \in \{0,1\}$, and as an \textit{Ising} formulation operating on $q_i \in \{-1,+1\}$. In both formulations, the form of the quadratic program \eqref{eq:quadratic_program} is unchanged.

As a convention, Qubo problems are often stated in the literature as a lower- (or upper-) diagonal matrix of weights $\left( Q_ij \right)_{i,j}$, where $Q_{ii}=a_i$ encode the linear weights and off-diagonal elements encode the quadratic couplers. Ising problems are sometimes given as a vector $h=(h_1,\ldots,h_n)=(a_1,\ldots,a_n)$ of linear weights and a lower- (or upper-) diagonal matrix $J_{ij} = Q_{ij}$ for $i<j$ ($i>j$) with couplers.

\subsection{Conversion of Ising to Qubo and vice versa}
\label{section_ising2qubo}
Qubo and Ising quadratic integer programs are equivalent. They only differ in the values the variables $q_i$ are allowed to take, and hence allow to be converted from one formulation to the other. This is done as follows.

Suppose we are given a Qubo problem with linear weights $Q_{ii}$, $i \in \{1,\ldots,n\}$, and quadratic weights $Q_{ij}$, $i<j$. Then, an equivalent Ising problem is given by setting
\begin{align*}
h_i &= \frac{Q_{ii}}{2} + \sum_{j=1}^n \frac{Q_{ij}}{4},\\
J_{ij} &= \frac{Q_{ij}}{4}
\end{align*}
for $i \in \{1,\ldots,n\}$ and all $i<j$.

Likewise, any Ising problem given by linear weight vector $h$ and coupler matrix $J$ can be converted to a Qubo by defining
\begin{align*}
Q_{ii} &= 2 \left( h_i - \sum_{j=1}^n J_{ij} \right),\\
Q_{ij} &= 4 J_{ij},
\end{align*}
for $i \in \{1,\ldots,n\}$ and all $i<j$.

\subsection{Qubit architecture and chains of qubits}
\label{section_architecture}
The D-Wave quantum computer operates on roughly $1000$ qubits arranged on a particular graph architecture. The precise number of available qubits varies from chip to chip due to manufacturing errors, meaning that slightly less of the envisaged qubits are usually operational in each device. The qubits are arranged in a \textit{chimera} architecture on the chip consisting of $4 \times 4$ bipartite graph cells, interconnected on a lattice of $12 \time 12$ cells.

The particular architecture of the qubits implies two important consequences: First, the chip design only allows for pairwise contributions of two qubits. Together with the linear contributions, this motivates the form of the objective function as a quadratic integer program in \eqref{eq:quadratic_program} which can be solved with D-Wave.

Second, although in theory, \eqref{eq:quadratic_program} includes pairwise interactions between any two qubits $q_i$ and $q_j$, this is not the case in practice. Since any physical qubit on the D-Wave chip possesses edges only within its $4 \times 4$ bipartite graph cell, as well as possibly some further connections to qubits of adjacent cells, pairwise interactions are actually very limited. Therefore, in practice, a pairwise interaction specified in the Qubo or Ising input problem has to be re-routed through various qubits in case the two qubits involved are not already neighbors on the physical chip. These additional qubits used to connect two bits $q_i$ and $q_j$ which are actually not neighbors on the physical chip are called a \textit{chain}.

The existence of chains has two vital consequences which will play an important role in Sections \ref{sec:maxc} and \ref{section_graph_partititioning}. On the one hand, the need for chains uses up qubits which would otherwise be available in the quadratic program, meaning that for problems containing many pairwise interactions, less than the around $1000$ qubits are available. In the worst-case of a complete graph, only $48$ qubits can be used in \eqref{eq:quadratic_program} when arbitrary pairwise couplers are desired. On the other hand, due to the probabilistic nature of the quantum annealing process, solutions returned by D-Wave are not always identical. This is a problem for all chains used in a particular instance, since all qubits in a chain encode the same single bit in \eqref{eq:quadratic_program} by construction, and should hence have a consistent value in any solution. This phenomenon is called a \textit{broken chain}. It is not clear which value a particular qubit shall be assigned if its chain is broken, making post-processing techniques necessary. Naturally, chains can be ensured to not break by assigning them higher coupler weights than in the actual quadratic program the user is interested in. In this case, however, the quantum annealer with be implicitly tuned towards ensuring that all chains are \textit{satisfied} (i.e.\ possess one consistent value), at the expense of violating some of the couplers of the actual quadratic program, thus leading to infeasible solutions. The trade-off between ensuring satisfied chains and putting sufficient weight on the actual quadratic program the user wishes to solve requires extensive further research.

\subsection{D-Wave solvers}
\label{section_dwave_solvers}
D-Wave Inc.\ provides several solvers to implement given instances of \eqref{eq:quadratic_program}, to submit them to the quantum computer, apply necessary post-processing steps to the solution it returns and to format the output. This section highlights the interfaces to the D-Wave computer used in this article.

\subsubsection{Sapi}
Sapi interfaces for the programming languages \textit{C++} and \textit{Python} are provided by D-Wave Inc. These interfaces allow to establish an internet connection to a D-Wave server, to submit quadratic programs in Ising and in Qubo format, and to post-process the solution.

Most importantly, the sapi interface allows the user to take own control over the D-Wave parameters: It allows to compute individual embeddings of the Ising or Qubo to the \textit{chimera} physical graph structure, to choose the number of anneals the quantum device is supposed to perform, or to specify the type of post-processing. The quadratic program itself is provided as a vector (containing linear weights) and matrix (containing quadratic weights) in appropriate data structures, solutions are returned as vectors which can be evaluated or processed further in any way the user sees fit. Although there are sapi routines for solving both Ising and Qubo formulations, we noticed that the Ising routines seem to work a lot better. This is not an issue since sapi routines to convert between the equivalent Ising and Qubo formulations are provided.

Since the user has full control over the graph embedding, quadratic programs can be provided which are designed to perfectly fit the \textit{chimera} architecture. Arbitrary graphs are limited to $48$ vertices (qubits), the maximal graph size that can be embedded onto the \textit{chimera} graph even when all pairwise interactions are present.

In summary, the sapi interface provides the lowest level control of the D-Wave system and is hence the best candidate for working directly on the quantum annealer.

\subsubsection{QBsolv}
QBsolv is a high-level tool that aims to take care of the D-Wave parameters and embedding issues automatically. It is not a programming interface but rather given by stand-alone binaries for selected computer architectures.

QBsolv solves Qubo problems only. The user provides a Qubo in a specific file format which lists linear weights first and quadratic couplers afterwards, where for each quadratic coupler $Q_{ij}$, the convention $i<j$ has to be satisfied.

Notably, instances submitted to QBsolv are not limited in size. They are automatically split up into subproblems submitted to D-Wave, and an extensive tabu search is applied to post-process all D-Wave solutions. QBsolv further allows to specify certain parameters such as the number of individual solution attempts, the subproblem size used to split up instances which do not fit onto D-Wave, a target value of the Hamiltonian that is sought or a timeout parameter.

The solution is either displayed in the terminal or written into an output file from which it has to be re-read for further processing.

\subsubsection{QSage}
QSage is another programming interface which allows to establish a connection to the D-Wave server within textit{C++} and \textit{Python} code. In contrast to Sapi or QBsolv, it is a blackbox solver which does not require a Qubo or Ising formulation.

Instead, QSage is able to minimize any one dimensional function operating on a binary input string of arbitrary size. To this end, QSage internally samples random points from the blackbox objective and uses the resulting function values to interpolate a quadratic program which is then submitted to D-Wave. The minimum returned by D-Wave is then used as a proxy for the minimum of the objective function and thus as a starting point for a tabu search or for further expansions of the objective function. Although QSage does not require the objective function to be smooth, naturally, best results are obtained in practice for smooth landscapes.

\subsection{Classical solvers}
\label{section_classical_solvers}
Apart from the tools provided by D-Wave Inc., we employ classical solvers in our comparison. These are described in this section.

\subsubsection{Simulated annealing with random starting point}
\label{section_classical_solvers_SA}
Simulated annealing (\textit{SA}) is a widely used optimization algorithm \citep{Kirkpatrick1983,DelMoral2006}. It provides an effective way to minimize (maximize) virtually any objective function using pointwise evaluations only.

SA iteratively minimizes an objective function by proposing a sequence of random moves (modifications) to an existing solution. These moves are drawn randomly according to a user-specified distribution. In order to give the algorithm the ability to escape local minima, modifications which do not further minimize the currently found best solution are not always rejected, but accepted with a certain \textit{acceptance probability}. This acceptance probability is proportional to the magnitude of the (unfavorable) increase in the objective function and antiproportional to the runtime. The latter gives SA the property to explore the objective function landscape at the beginning of a run while gradually turning into a steepest descent algoritm over time. In order to make the acceptance probability antiproportionally depend on the runtime, a so-called \textit{temperature} function is employed. This is any decreasing function to zero. SA can be run with a variety of stopping criteria, for instance with a fixed number of iterations (as in our case), until no better solution has been found for a certain number of iterations, or when the temperature becomes vanishingly small.

Algorithm \ref{algorithm_sa} shows an implementation of SA adapted from \cite{KadowakiNishimori1998} for solving graph partitioning instances given by an Ising Hamiltonian (see Section \ref{section_hamiltonians_ising}). It uses the common temperature function $T(t) = \log(1+t)$ \citep{GemanGeman1984}.
\begin{algorithm}[t]
\caption{\texttt{Simulated Annealing for Ising model}}
\label{algorithm_sa}
\SetKwInOut{Input}{input}
\SetKwInOut{Output}{output}
\Input{Ising problem in $n$ variables $q_i \in \{+1,-1\}$, number of steps $t_{\max}$,
initial random state with $|\{q_i:q_i=+1\}|=|\{q_i:q_i=+1\}|=\frac{1}{2}|V|$}
\Output{$q_1^\ast,\ldots,q_n^\ast$ of local minimum}
\For{$t \leftarrow 1$ \KwTo $t_{\max}$}{
  Evaluate Ising model on current $q_i$ and store energy as $E$\;
  Select one random bit each among $\{q_i:q_i=+1\}$ and $\{q_i:q_i=-1\}$\;
  Flip value of both bits and recompute Ising energy as $E'$\;
  $\Delta \leftarrow E'-E$\;
  $p \leftarrow \exp \left(-\Delta/T(t) \right)$\;
  Keep both bits with probability $\min(1,p)$, otherwise flip both back\;
}
Return $q_i^\ast=q_i$, $i \in \{1,\ldots,n\}$\;
\end{algorithm}

\subsubsection{METIS and Simulated Annealing}
\label{section_classical_solvers_METIS_SA}
\texttt{METIS} \citep{KarypisKumar1999} is a popular heuristic multilevel algorithm to perform graph partitioning based on a three-phase approach:
\begin{enumerate}
  \item The input graph is coarsened by generating a sequence $G_0, G_1, \ldots, G_n$ of graphs starting from the original graph $G=G_0$ and ending with a suitably small graph $G_n$ (typically less than 100 vertices). The sequence has the the property that $G_i$ contains less nodes than $G_j$ whenever $i<j$. To coarsen the graph, several vertices are combined during the coarsening phase into a \textit{multinode} of the next graph in the sequence.
  \item $G_n$ is partitioned using some other algorithm of choice, for instance spectral bisection or $k$-way partitioning.
  \item The partition is projected back from $G_n$ to $G_0$ through $G_{n-1},\dots,G_1$. As each of the finer graphs during the uncoarsening phase contains more degrees of freedom than the multinode graph, a refinement algorithm such as the one of Fiduccia-Mattheyses \citep{FiducciaMattheyses1982} is used to enhance the partitioning after each projection.
\end{enumerate}
\texttt{METIS} allows the user to change several parameters, for instance the size of the subgraphs $G_n$, the coarsening or the partitioning algorithms.

We employ \texttt{METIS} in two ways. First, to calculate edge-cut partitionings of given graphs (Sections \ref{section_hamiltonians_ising} and \ref{section_hamiltonians_qubo}) and to use its partitions as a starting point for \textit{core-halo} (CH-) partitioning \cite{Djidjev2016} investigated in Section \ref{section_hamiltonians_ch}.

For CH-partitioning, \cite{Djidjev2016} provide an algorithm based on Simulated Annealing capable of refining a given partitioning in order to optimize it further with respect to a different metric. Whenever a CH-partitioning is sought in Section \ref{section_graph_partititioning}, we thus employ the Simulated Annealing approach of \cite{Djidjev2016} with initial solution provided by \texttt{METIS}.

\subsubsection{Gurobi}
\label{section_classical_solvers_Gurobi}
\texttt{Gurobi} \citep{gurobi} is a mathematical programming solver for linear programs, mixed-integer linear and quadratic programs, as well as certain quadratic programs.

We employ \texttt{Gurobi} to solve given Qubo problems (Ising problems can be solved as well, nevertheless \texttt{Gurobi} explicitly allows to restrict the range of variables to binary inputs, making it particularly suitably for Qubo instances). To this end, we programmed a simple converter which takes a given Qubo and outputs a quadratic program in \texttt{Gurobi} format.

\section{The maximum clique problem}
\label{sec:maxc}
\newcommand{\DW}{D-Wave 2X}

\subsection{Introduction}
The maximum clique problem is a famous NP-Hard problem, with applications in social networks, bioinformatics, data mining, and many other fields. Its formulation is quite simple, given an undirected graph  $G = (V, E)$, a clique is a subset $S$ of the vertices forming a complete subgraph, i.e. where every two vertices of $S$ are connected by an edge in $G$. The clique size is the number of vertices in $S$, and the maximum clique problem is to find the largest possible clique in $G$. 

A previous study investigated the use of quantum adiabatic evolution to solve the clique problem \cite{childs2000finding}. However, the \DW system was not available at that time and the study was therefore limited to small graphs ($n \leq 18$). Its conclusion was that their quantum algorithm appeared to require only a quadratic run time. Our objective is to verify that assumption on the real \DW system, on much larger graphs if possible.

\subsection{QUBO formulation}
A QUBO is written as 
$$\sum_{i\geq j} Q_{ij}x_{i}x_{j}$$
with  $x_{i} \in \{0,1\} $ and $i \in [0;N]$. The $Q_{ij}$ with $i \neq j$ are the quadratic terms, and $Q_{ii}$ are the linear terms. Paper from A.Lucas provides this QUBO formulation of the clique problem \cite{Lucas2014}:
$$H =  A \left( K- \sum_v x_v \right)^2 + B \left[ \frac{K(K-1)}{2} - \sum_{(uv) \in E} x_ux_v\right].$$

The ground state will be $H=0$ when a clique of size $K$ exists. Extra variables are then added to find the largest clique in the graph. However, this formulation requires $N=|V|$ variables on a complete graph, because of the $( K- \sum_v x_v )^2$ term that will generate every posible quadratic term in the QUBO formula. We prefer a simpler formulation, derived from the \emph{maximum independent set} (MIS) problem.

An independent set of a graph $G$ is a set of vertices, where every two vertices are not connected by an edge in $G$. It is easy to see that an independent set of $G = (V, E)$ is a clique in the complement graph $H = (V, \overline{E})$, where $\overline{E}$ is the set of edges not present in $G$. Therefore looking for the maximum clique in $G$ or the maximum independent set in $H$ is the same problem. The QUBO formulation of the MIS problem on a graph $H$ is straightforward, let $x_i=1$ if vertex $i$ is in the independent set, and $x_i=0$ otherwise. We should have a penalty when two vertices in this set are connected by an edge in the graph, therefore we need some positive quadratic term $M$ over all existing edges of the graph $H$. We want the largest independent set, therefore a negative weigth L on the linear terms. The penalty for an edge should be greater than the bonus for a vertex, so that it is detrimental to add a vertex with a single edge to the set, so for example $L=-1$ and $M=2$.

The QUBO for the MIS problem of a graph $G = (V, E)$ is therefore:
$$-\sum_{i \in V} x_i + 2\sum_{(i,j) \in E} x_ix_j.$$

This formulation requires $|V|$ variables and $|E|$ quadratic terms. Note that the ising graph (the set of edges and vertices with non zero weigth, also called the primal graph) required for this problem is the graph $G$ itself.

When solving the maximum clique problem for a graph $G_1 = (V,E)$ we need to complement the graph first to $G_2= (V, \overline E)$ and solve the MIS on $G_2$, thus requiring  $|V|$ variables and $|\overline E| =  \frac{|V| (|V|-1)}{2} - |E| $ quadratic terms. It will therefore be easier to solve the maximum clique problem on a very dense graph, that will generate a sparse complement graph.

\subsection{Methods}
The \DW system consists in a specific graph topology called the chimera graph. It is a lattice of 12x12 cells, where each cell is a 4x4 bipartite graph. \DW can naturally solve ising problems where non-zero quadratic terms are represented by an edge in the chimera graph. For other problems, an \emph{embedding} is needed, \emph{i.e} a logical variable has to be represented by a chain of physical qubits, in order to increase the connectivity of the simulated variable. The largest complete graph that the \DW can embed in theory has $1+4*12 = 49$ vertices. It is slightly smaller ($n=45$) because of missing qubits coming from manufacturing defaults.

In our problem, the maximum number of connections needed may be up to a complete graph. The largest graph that we are sure to be able to solve on \DW is therefore a 45-vertex graph. For larger graphs, it will depend on the edge density and topology.

To solve the MIS problem on any graph, we therefore require a divide-and-conquer strategy that first splits the input graph into smaller graphs of at most 45 vertices.

\subsubsection{Sapi parameters}
Sapi stands for solver API, it is the lowest level of control we can have over the quantum annealer. We use here the sapi C client version 2.4.2. We first convert our QUBO problem to an ising problem. We use the pre-computed embedding of the full 45-vertex graph for the experiments in section~\ref{sec:anyg}. In section~\ref{sec:anyg} we do not need to compute an embedding. We use num\_reads=500, postprocessing=SAPI\_POSTPROCESS\_OPTIMIZATION and provide the embedding to the ``chains'' parameters of the solver so that the post-processing is applied to the original problem without chains (with this parameter the solver first applies majority voting to resolve chains, before applying post-processing). Our experiments showed that this is the best way to deal with chains.

\subsection{Results}

\subsubsection{Software solvers}
Results from \DW are compared to different solvers. A simulated annealing algorithm working on the ising problem (SA-ising), a simulated annealing algorithm specifically designed to solve the clique problem (SA-clique \cite{geng2007simple}), a software designed to find cliques in heuristic or exact mode (fmc \cite{pattabiraman2013fast}), and the gurobi solver. We also include in the comparison results coming from the Dwave post-processing heuristics only.

\paragraph{SA-ising}
This is a simulated annealing algorithm working on an ising problem. The intial solution is a random solution, and a single move in the simulated algorihtm is the flip of one random bit.

\paragraph{SA-clique}
We implememted the simulated algorithm designed to find cliques described by X.Gen \emph{et al.} \cite{geng2007simple}. It is designed to find a clique of a specific size $m$, we therefore need to apply a binary search on top of it to find the maximum clique size. The main parameter is the $\alpha$ value controlling the temperature update at each step: $T_{n+1} = \alpha T_n$. They typically use $\alpha=0.9996$. A value closer to 1 will yield a better solution but will increase computation time.

\paragraph{Fast Max-Clique Finder (fmc)}
This algorithm is designed to find efficiently the maximum clique for a large sparse graph. It provides an exact and a heuristic search mode. We use the 1.1 version of the fmc software \cite{pattabiraman2013fast}.

\paragraph{Post-processing heuristics alone (PPHa)}
The Dwave pipeline includes a post-processing step: first, if chains exist, a majority vote is applied to fix the broken chains. Then a local search is performed so that solutions become locally minimum (the raw solutions coming from dwave might not be locally minimum) \cite{dwavepostprocess24}. For a given solution coming out of the pipeline, one might wonder what are the relative contributions of the \DW system and of the post-processing step. For some small and simple problems, the post processing step \emph{alone} might be able to find a good solution.

We try to answer this by applying post-processing step only, and see how it compares with the actual quantum annealing. However, this post-processing step runs on the dwave server and is not available separately. 

We therefore need some special procedure: we set a very high absolute chain strength (e.g. 1000 times greater than the largest weight in our ising problem), and turns on the ``auto-scale'' feature. Consequently, chains weights will be set to minimum value -1 and all other weights will be scaled down to 0. In this way, the quantum annealer will only satisfy the chains, and will not solve anything else. Each chain will not be connected to anything, and all linear terms will be zero, therefore each chain should get a random value $-1$ or $+1$. The post-processing step will therefore be fed with some random initial solution, and what we get at the end of the pipeline should be the result of post-processing step only. This will be referred to later as PPHa, \emph{post-processing heuristic alone}.

\subsubsection{Results}

\paragraph{Results on a small graph}
We first analyze the behavior of the quantum annealer on one graph of size at most 45 vertices, i.e. the largest graph we are sure to be able to embed. \label{sec:anyg}

\begin{table}
\centering
\begin{tabular}{lllllll}
\hline
Graph & Max. Clique size & Sapi & PPha & QBsolv & fmc & SA\\
\hline
Random 0.3 &  5   &  0.15 s & 0.15 s & 0.05 s & $8.10^{-6}$ s & 0.15 s\\
Random 0.5 &  8   &  0.15 s & 0.15 s & 0.06 s & $3.10^{-4}$ s & 0.37 s\\ 
Random 0.7 &  13 &  0.15 s & 0.15 s & 0.04 s & 0.002 s & 0.19 s\\
Random 0.9 &  20 &  0.15 s & 0.15 s & 0.04 s & 0.135 s & 0.28 s\\
\hline
\end{tabular}
\caption{ Running time on 45 vertices random graph. Graph is generated with probability of edge present equal to number in first column. For these small graphs, every software return the correct solution, therefore only the running time is reported.}
\end{table}

\begin{figure}
\centering
\includegraphics[width=0.5\textwidth]{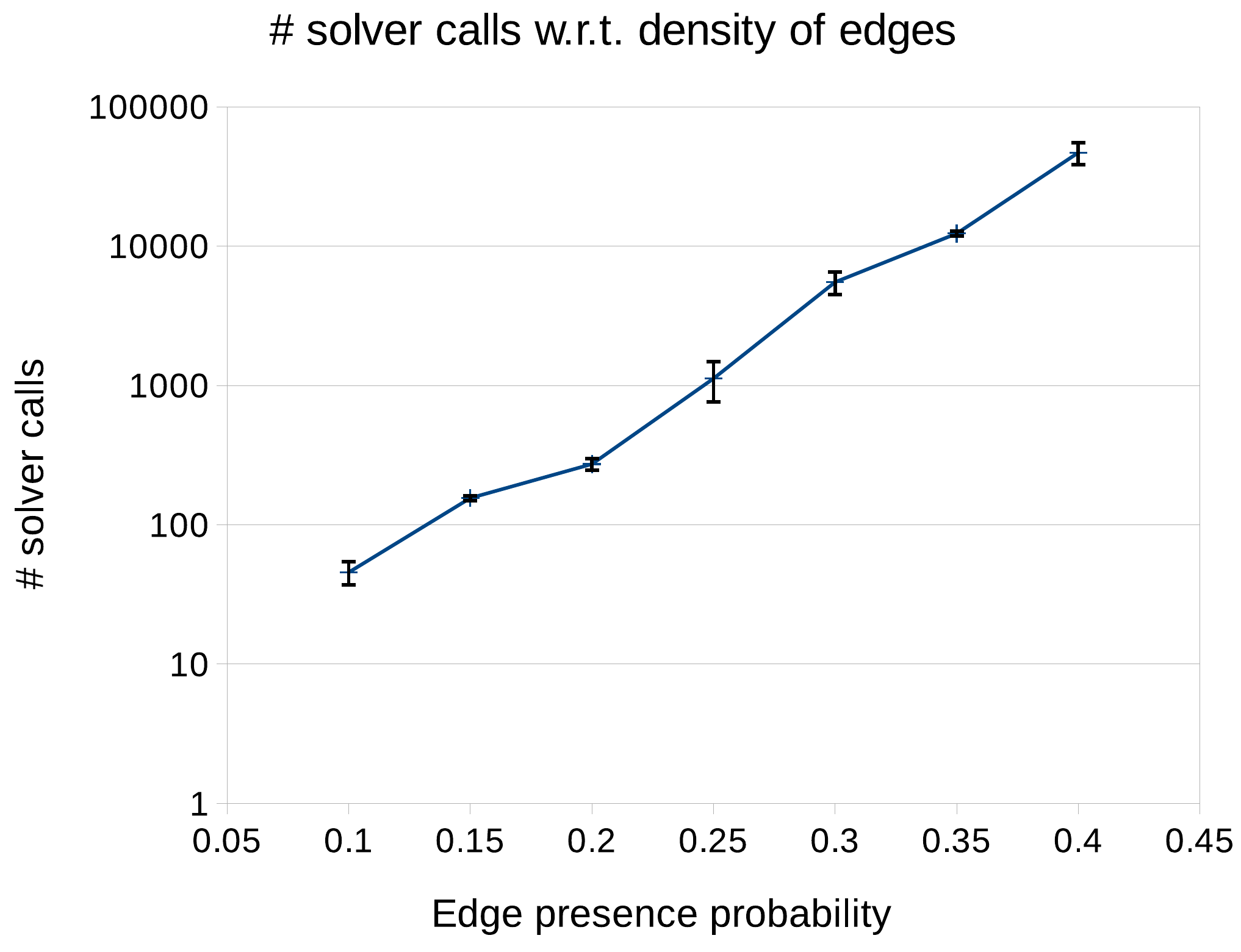}
\protect\caption{The number of solver calls follows an exponential trend with increasing graph density.\label{fig:Number-of-solver}}
\end{figure}

We test our graph splitting method on random graphs with 500 vertices and increasing density. The density of the generated graph is controlled by the probability of the presence of an edge ranging from 0.1 to 0.4. Fig.~\ref{fig:Number-of-solver} shows the results of this experiment. We report here the number of solver calls required to solve the input graph -ie. the number of generated subgraphs- with respect to the edge presence probability. Each data point is the median value of ten runs with the standard deviation given as error bars. The number of solver calls follows an exponential curve with respect to the density of the input graph.

\begin{figure}
\centering
\includegraphics[width=0.5\textwidth]{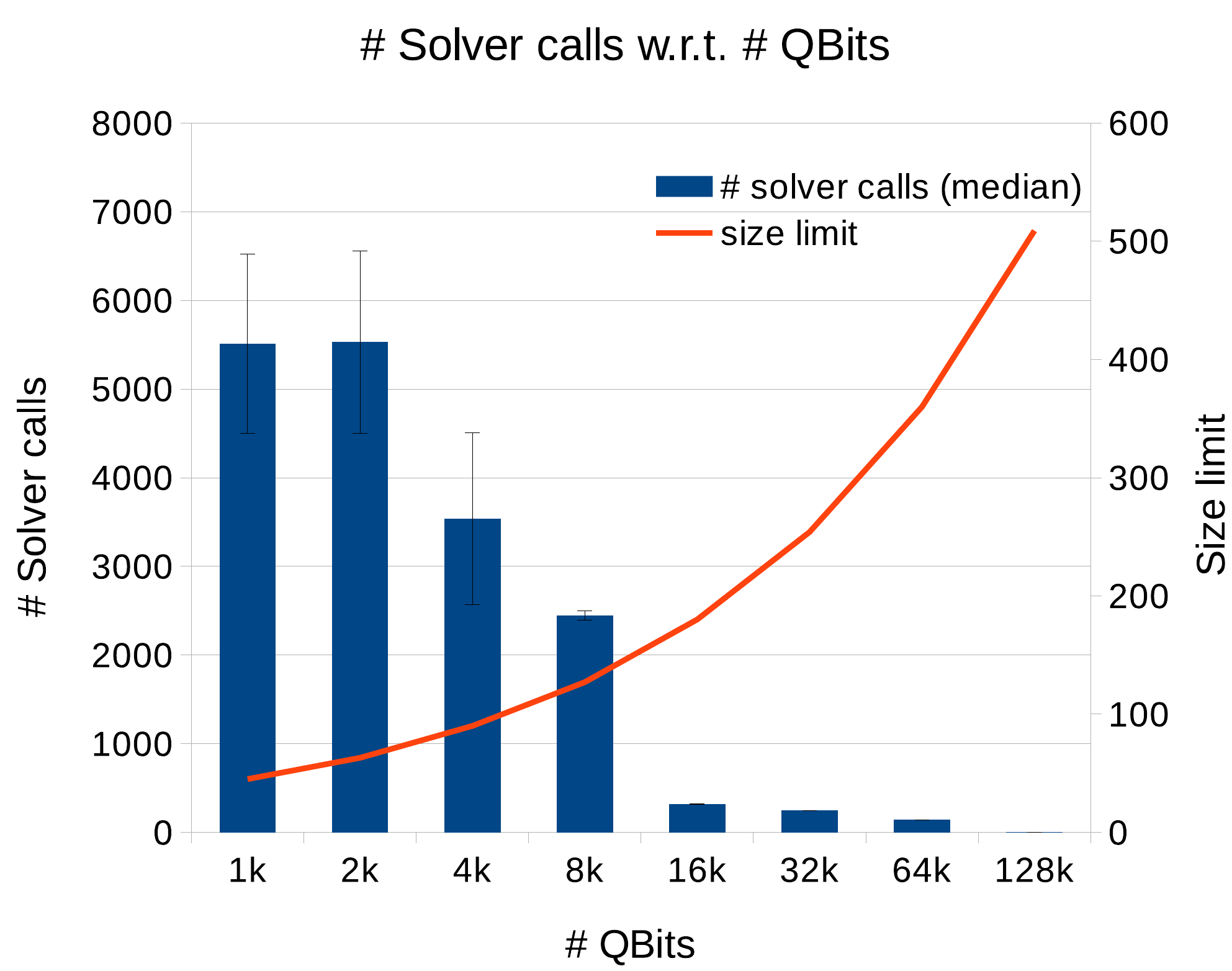}
\protect\caption{Number of solver calls (left y-axis) with increasing number of QBits. The size limit - ie. the maximal arbitrary graph that can be embedded on a given DWave system - can be read on the right y-axis.\label{fig:Number-of-solver-1}}
\end{figure}

In a second experiment, we measure the impact of future generations of DWave systems on our clique finding approach for arbitrary large graphs. Assuming a similar chimera topology for future generations of DWave systems, doubling the number of available QBits will increase the size of the maximal complete subgraph that can be embedded by a factor $\sqrt{2}$. Fig~\ref{fig:Number-of-solver-1} shows the evolution of the number of solver calls for increasing number of QBits. In this experiment the graph size is fixed at 500 vertices and the edge presence probability at 0.3. Each data point is the median of ten runs with standard deviation given as error bars. If we consider that the number of QBits doubles with each new generation, 7 generations of DWave machine are required to directly embed and solve an arbitrary 500 vertex graph.

\subsubsection{Artificial graph designed to fit chimera}
\label{sec:artig}
The idea here is to try to see what is the speedup \DW can achieve in the best possible scenario. This best possible scenario is a graph that fits well the chimera graph, i.e. a graph that requires a very few chains, and hopefully for which the MIS problem is not trivial.

\paragraph{Graph generation}
We start with the actual chimera graph $C = (V,E)$ of the \DW system, in our case $|V = 1100|$  qubits (this may vary from system to system because of missing qubits).  This is the most dense 1100-vertex graph the system can solve. Note that this graph is not interesting for the MIS problem since it is bipartite, and one could argue that MIS on a bipartite graph is a trivial problem.

Consider now the graph $C_1$ obtained by contracting one random edge from $C$. An edge contraction is the removal of an edge $v_1,v_2$ and fusion of vertices $v_1$ amd $v_2$.  With $\mathcal{N}_1$ and $\mathcal{N}_2$ the set of neighbouring vertices from $v_1$ and $v_2$, the neighbours of the fused node are $\mathcal{N}_1\cup   \mathcal{N}_2 \setminus \{v_1,v_2\} $. $C_1$ has $|V|-1$ vertices and is called a \emph{minor} of the chimera graph. By definition the ising formulation of the MIS problem\footnote{The primal graph of the ising MIS problem of a graph is the graph itself.} for this graph can be embedded: the embedding is given by the edge contraction. Moreover, if we add any edge to $C_1$, the resulting graph will not be embeddable into chimera, since $C_1$ already uses every possible qubits and edges of the chimera graph. We can therefore say that $C_1$ is one of the most dense graph of size $|V| -1$ than chimera can embed.

We can generalize after $m$ random edge contractions: the resulting graph $C_m$ will have $|V| -m$ vertices, will be one of the most dense graph of size $|V| -m$ chimera can embed, and will not be bipartite. This family of graphs $C_m$ with $0<m<1100$ is therefore a good candidate for the best case scenario experiment for the MIS problem: large graphs that can embed on chimera, and for which solution is not trivial. 
 
\begin{figure}
  \centering
  \includegraphics[width=0.5\textwidth]{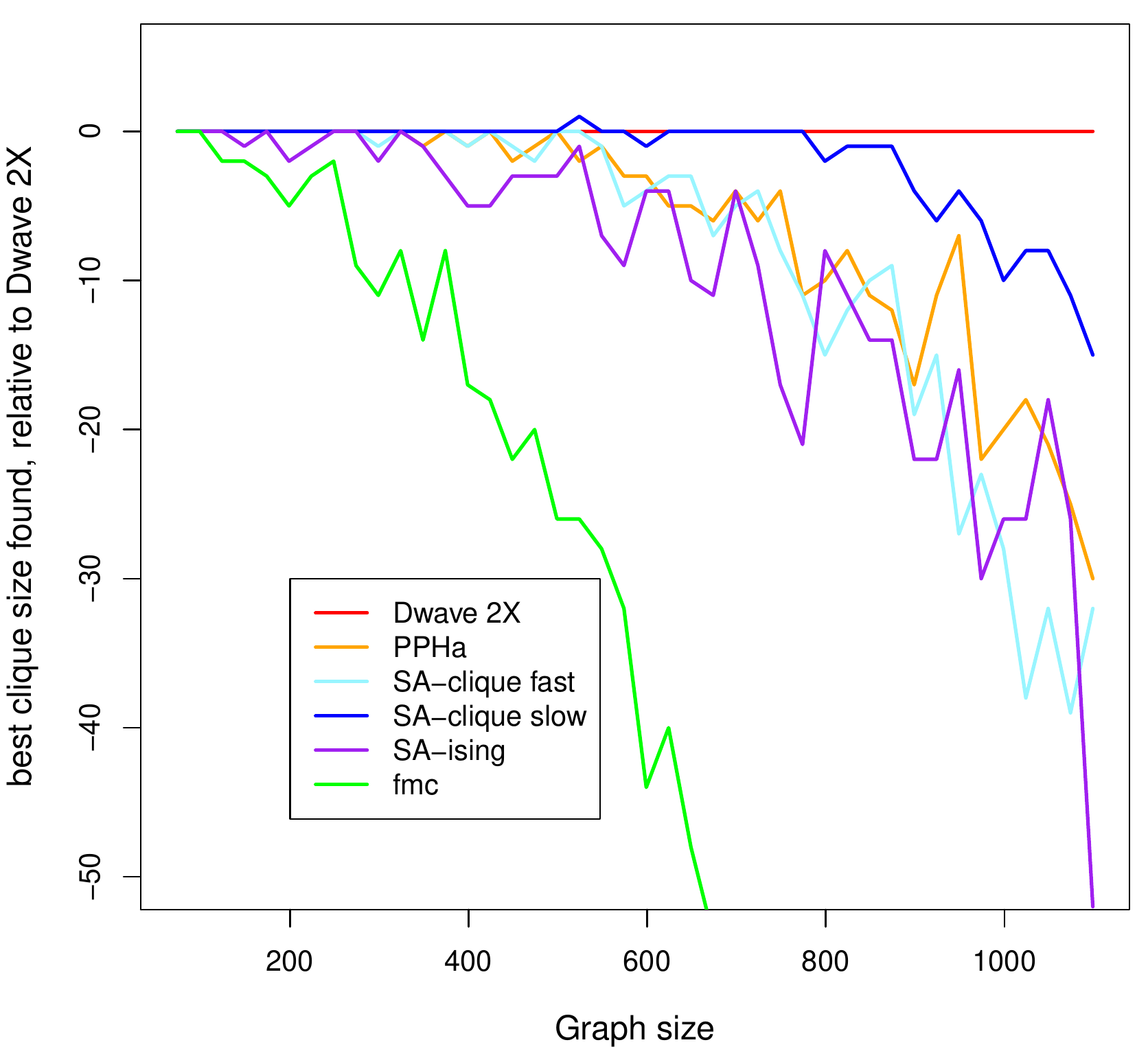}
  \caption{Best clique size found by the different solvers, relatively to the \DW result, on the  $C_m$ family of graphs. These are articial graphs designed to fit the chimera topology, representing a best case scenario for Dwave. $PPHa$ stands for \emph{post-processing heuristics alone}. SA-clique fast and SA-clique slow have $\alpha$ parameters $0.9996$ and $0.99999$. SA-ising has $\alpha=0.9999$. $fmc$ is running in its heuristic mode.}
  \label{fig:Cm}
\end{figure}

\paragraph{Experiment}
We solve the MIS problem on the $C_m$ family of graphs, using Sapi on the \DW system, $PPHa$ and the $SA-ising$ software. We solve the equivalent maximum clique problem on the complement graphs using $SA-clique$ and $fmc$. Figure~\ref{fig:Cm} shows the result.  We can see that for graph sizes up to 400, $PPHa$ gets same result as Dwave. For these small graphs the problem is probably simple enough to be solved by the post-processing step alone. As expected, the simulated annealing designed specifically for clique is behaving better than SA-ising. $fmc$ software is run in its heuristic mode. It is designed for large sparse graphs, and run here on very dense graphs, probably the reason of comparatively lower quality result. 

For large graphs (larger than 800 vertices), \DW gives the best solution. Note that we do not know if that is the optimal solution.

\begin{figure}
  \centering
  \includegraphics[width=0.5\textwidth]{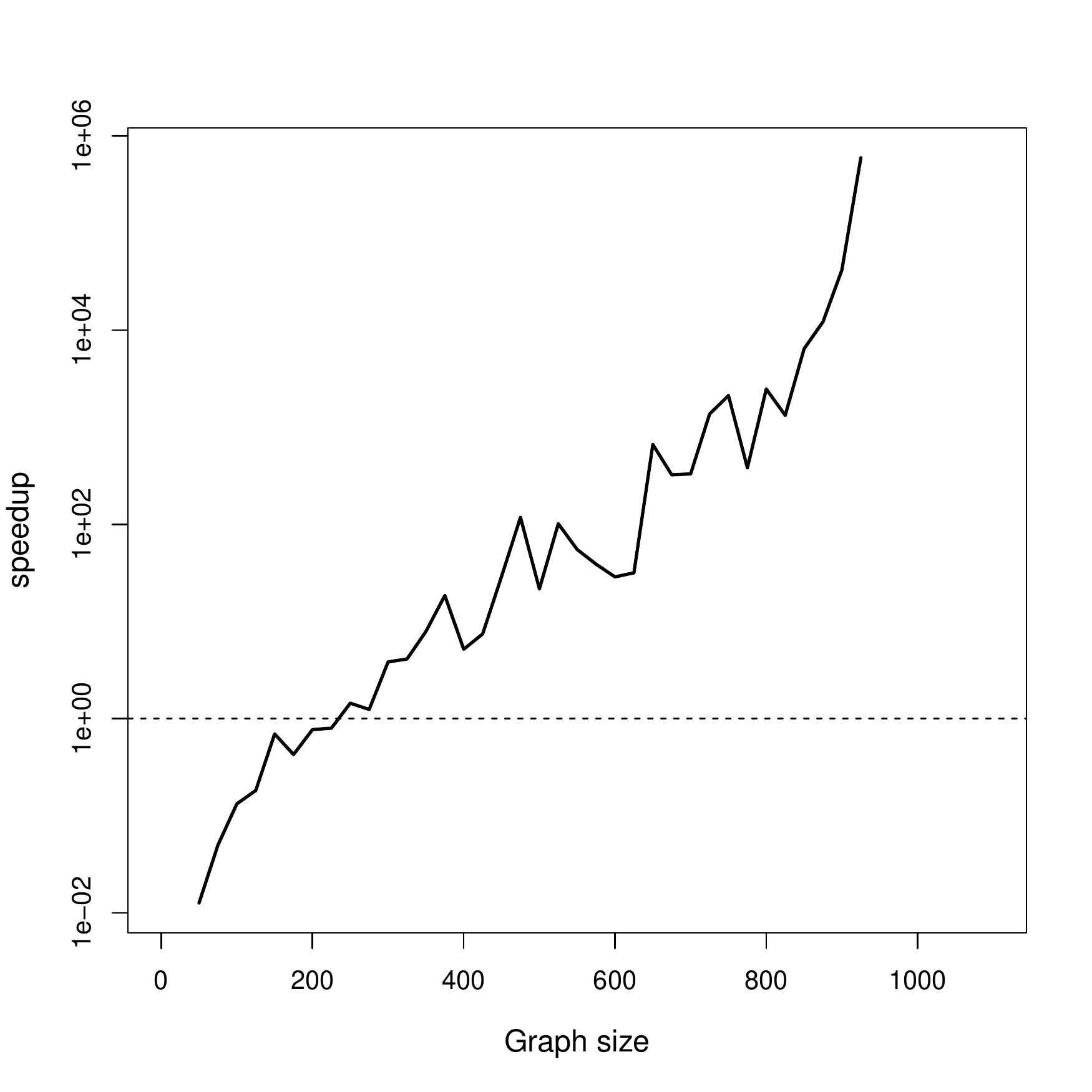}
  \caption{Speedup on artificial graphs designed to fit the chimera topology. \DW speedup is computed versus a simulated annealing algorithm designed for cliques (SA-clique), running on one cpu core of an Intel E8400 @ 3.00GHz. For SA-clique, we progressively increase the cooling schedule (i.e. increasing running times), and report the fastest running time for which SA-clique gets same result as \DW. For small graphs, the simulated annealing algorithm is able to find very quickly the same result as \DW. For large graphs, it becomes exponentially more difficult,  the SA algorithm requires a very slow cooling schedule to get the target solution, leading to a \DW speedup over a million. Timings are averaged over multiple runs. }
  \label{fig:speedup}
\end{figure}

\paragraph{Speedup}
Since SA-clique seems to best the best candidate to compete with \DW, we choose to compute the \DW speedup relatively to SA-clique on this $C_m$ family of graphs. The procedure is as follows: for each graph size, we run \DW with 500 anneals and report the best solution. The \DW runtime is the total qpu runtime for 500 anneals, i.e. approximately $0.15s$. For SA-clique, we start with a low $\alpha$ parameter (i.e. a fast cooling schedule), and gradually increase $alpha$ until SA-clique finds the same solution as \DW. This $alpha$ for which SA-clique finds the same solution gives us the SA-clique best execution time. The SA-clique algorithm is run on one CPU core of an Intel E8400 @ 3.00GHz.  Figure~\ref{fig:speedup} shows the speedup for different graph sizes (of the $C_m$ family). For graph sizes lower than 200 vertices, \DW is slower, but for larger grah sizes, it gets exponentially faster, up to a million speedup. 

Although this is the best possible case scenario for \DW and not representative of what we would get with random graphs, this may be an evidence showing quantum behavior in the \DW system.  \DW is able to find very quickly a solution that is \emph{very} difficult to obtain with all the classical solvers we tried.

\section{Graph partitioning}
\label{section_graph_partititioning}
This section investigates a variety of methods for graph partitioning. Our focus is on two variations of the graph partitioning problem: Edge-cut partitioning and \textit{core-halo} partitioning \citep{Djidjev2016}. To this end, Section \ref{section_hamiltonians} provides Hamiltonians in either Ising or Qubo formulation to carry out both types of partitioning. We then employ the methods outlined in Section \ref{section_methods} to minimize those Hamiltonians on a variety of graphs in order to assess their performance in a simulation study (Section \ref{section_simulations}).

\subsection{Hamiltonian formulations}
\label{section_hamiltonians}
\subsubsection{Ising model for graph partitioning}
\label{section_hamiltonians_ising}
\cite{Lucas2014} gives an Ising Hamiltonian to carry out standard edge-cut partitioning. However, this formulation is only suitable to partition a graph into two components. A generalisation to $n$ components will be derived in the next section.

Edge-cut partitioning aims to divide the graph into partitions in such a way that the edge-cut, the number of edges connecting vertices belonging to different partitions, is minimized.

To this end, \cite{Lucas2014} considers an Ising spin $s_v \in \{ +1,-1\}$ per vertex $v \in V$ indicating which of the two partitions $v$ belongs to (one being the "$+$" partition and one being the "$-$" partition). He defines the Hamiltonian as $H=A H_A + B H_B$, where
\begin{align*}
H_A &= \left( \sum_{i=1}^n s_i \right)^2,\\
H_B &= \sum_{(u,v) \in E} \frac{1-s_u s_v}{2}.
\end{align*}
In the above, $H_A$ is a penalty term enforcing equal balance of vertices in both partitions. Assuming an even number of vertices, $H_A$ is zero if and only if the number of vertices in both partitions is equal. For an odd number of vertices, $H_A \geq 1$. The term $H_B$ is the objective function. It adds a penalty of $1$ for any edge connecting two vertices which are not in the same partition and is thus equal to the edge-cut. Minimizing $H$ will thus lead to a minimal edge-cut partitioning of the graph.

The weights $A$ and $B$ have to be chosen dependent on the input graph in such a way that it is never favorable to violate $H_A$ in order to achieve a further decrease in the objective function $H_B$. To this end, it suffices to determine the minimal increase in the penalty for any given change while maximizing the decrease in the objective function. Consider the configuration of $s_v$, $v \in V$, at the global minimum (the equilibrium state) of $H$. For simplicity, assume without loss of generality that the number of vertices is equal and hence that $H_A=0$.

When changing the spin of one $s_v$, the change in $H_A$ is $\Delta H_A \geq 4$ since $s_v$ was previously compensated in equilibrium by another spin indicator. We now consider any possible decrease in $H_B$. In the best case, all the neighboring vertices of $v$ did not belong to the partition $v$ was in since in this case, a change in $s_v$ will cause $\frac{1-s_u s_v}{2}$ to change from one to zero for all $w \in N(v)$, where $N(v)$ is the set of neighboring vertices of $v$. Hence, the decrease in $H_B$ can be bounded above by $\Delta H_B \leq \Delta$, where $\Delta$ denotes the maximal degree of any vertex in $G$. To prevent an increase in $H_A$ to be ever favorable in order to achieve a decrease in $H_B$, we must set $A \Delta H_A \geq 4A > \Delta B \geq B \Delta H_B$, thus $A > \Delta/4 B$. Setting $B=1$ leads to the choice $A = \Delta/4+1$.

\subsubsection{A generalisation to partitioning into multiple parts}
\label{section_hamiltonians_qubo}
The Hamiltonian considered in Section \ref{section_hamiltonians_ising} can easily be generalised to an arbitrary number of partitions. For this, let $s_{vi} \in \{ 0,1 \}$ be a binary (qubo) indicator which encodes if vertex $v$ belongs to partition $i$. Let $K$ be the desired number of partitions.

Define $H=A H_A + B H_B + C H_C$, where
\begin{align*}
H_A &= \sum_{v \in V} \left( \sum_{k=1}^K s_{vk} -1 \right)^2,\\
H_B &= \sum_{k=1}^K \left( \sum_{v \in V} s_{vk} -\frac{|V|}{K} \right)^2,\\
H_C &= \sum_{(u,v) \in E} \sum_{k=1}^K (1-s_{uk} s_{vk}).
\end{align*}
The reasoning behind the above formulation is as follows. $H_A$ is zero if and only if each vertex $v$ only belongs to exactly one partition, that is if for each $v \in V$ there exists only one $k_v$ such that $s_{v,k_v}=1$. Since $\sum_{v \in V} s_{vk}$ is the size of partition $k \in \{1,\ldots,K\}$, $H_B$ enforces partitionings of equal size $|V|/K$. The term $H_C$ is the objective function: It incentivises two vertices $u$ and $v$ to belong to the same partition, that is $s_{uk}=s_{vk}=1$ (for partition $k$, say), and thus minimizes the edge-cut between partitions.

As in Section \ref{section_hamiltonians_ising}, the weights $A$ and $B$ for the two penalty terms and the weight $C$ for the objective function have to be chosen appropriately. Consider $H$ in a global minimum and an arbitrary $s_{vk}$. Suppose first that $s_{vk}=1$ in the optimal solution and hence changes to $0$. Since we assume optimality, $H_A=H_B=0$, meaning that a change of $s_{vk}$ from $1$ to $0$ results in $\Delta H_A \geq 1$, $\Delta H_B \geq 1$ and $\Delta H_C \geq 0$ (in $H_C$, a change of any $s_{vk}$ from $1$ to $0$ potentially only results in terms $1-s_{uk} s_{vk}$ switching from $0$ to $1$). This means that changing an indicator $s_{vk}$ with value $1$ in the optimal solution is never favorable to achieve a further decrease in the objective function.

Next, assume that $s_{vk}$ changes from $0$ to $1$. In this case, $\Delta H_A \geq 1$ and $\Delta H_B \geq 1$ as before. However, a change of any $s_{vk}$ from $1$ to $0$ potentially cancels up to $\Delta$ (the maximal degree of the graph) terms of the form $1-s_{uk} s_{vk}$ in $H_C$, hence resulting in a decrease of the objective function of up to $\Delta H_C \leq \Delta \leq |V|$. To prevent that violating the penalty terms $H_A$ and $H_B$ ever leads to a decrease in $H_C$, we must hence ensure that $A \Delta H_A+B \Delta H_B \geq A(1+1) > C |V| \geq C \Delta H_C$, meaning that $A+B > C |V|$. Setting $C=1$ leads to the condition $A+B>|V|$. Since the constraints lead to an under-determined equation, more than one choice for $A$ and $B$ exists. For simplicity, splitting up the weight equally among both penalty terms (that is, $A=B$) leads to the choice $A=B=\frac{1}{2}|V| +1$.

\subsubsection{Qubo model for CH-partitioning}
\label{section_hamiltonians_ch}
Apart from standard edge-cut partitioning, we consider a special flavor of the graph partitioning problem termed \textit{core-halo} (CH) partitioning in \citep{Djidjev2016}. In CH-partitioning, we aim to find a partitioning of the vertices $V$ into $K$ sets (partitions) in such a way that the (squared) sum of both the number of vertices $c_i$ (the \textit{core}) in each partition $i \in \{ 1,\ldots,K \}$ as well as the number of neighboring vertices in other partitions $h_i$ (the \textit{halo}) is minimized:
\begin{align}
\min_{(c_i,h_i)_{i=1,\ldots,K}} \sum_{i=1}^K (c_i+h_i)^2.
\label{eq:sum_of_squares}
\end{align}

In order to design a Hamiltonian for this problem consider $H=A H_A + B H_B + C H_C$, where
\begin{align*}
H_A &= \sum_{v \in V} \left( \sum_{k=1}^K c_{vk} -1 \right)^2,\\
H_B &= \sum_{v \in V} \sum_{i=1}^K \left( \sum_{w \in N(v) \cup \{v\}} (h_{vi} - c_{wi} - z_{(v,w),i})^2 \right),\\
H_C &= \sum_{i=1}^K \left( \sum_{v \in V} h_{vi} \right)^2
\end{align*}
and $N(v)$ denotes the set of all neighbors of $v \in V$. We use two types of indicators to encode whether a vertex is in the core or in the halo of a given partition. Each variable $c_{vi}$ is a binary (qubo) indicator encoding with a value of one that vertex $v$ belongs to partition (core) $i$. Likewise for the halo indicators $h_{vi}$. The auxiliary variables $z_{(v,w),i}$ for each edge $(v,w)$ in the graph and each partition $i$ have no interpretation. They are used to ensure that the penalty $H_B$ takes the value zero in equilibrium.

The reasoning behind the above formulation is as follows. As each vertex can only belong to one core, as in Section \ref{section_hamiltonians_qubo}, only one of the indicators $c_{v1},\ldots,c_{vK}$ can take the value one for each $v \in V$. This is enforced in $H_A$ which is zero if and only if there is exactly one $i$ such that $c_{vi}=1$ for each $v \in V$.

Moreover, by definition of the CH-partitioning problem, a vertex $v \in V$ belong to the halo of a partition $j$ if and only if there is an edge in the graph connecting $v$ to a vertex $w$ in the core of partition $j$. We extend this definition and define $v$ to be a halo vertex of partition $j$ if either it is part of core $j$ or if there is an edge connecting $v$ to a vertex in core $j$. This is enforced in $H_B$: Given a $v \in V$ and partition number $i$ (see the two sums in $H_B$), $h_{vi}$ is forced to be one whenever either $c_{vi}=1$ or $c_{wi}=1$ for any $w \in N(v)$, the set of all neighbors of $v$. Naturally, one neighbor $w$ (of $v$) in partition $i$ suffices to cause $h_{vi}=1$. Therefore, $h_{vi}$ can be one even if $c_{wi}=0$ for a given $w$, thus causing the penalty to not attain the value zero in equilibrium. We correct this by introducing variables $z_{(v,w),i}$ which in the latter case will take the value $1$ and ensure that $H_B=0$ in equilibrium. If $h_{vi} - c_{wi}=0$ already holds true for given $v$ and $w$, there is no incentive for $z_{(v,w),i}$ to be one since this increases the overall value of the Hamiltonian.

The objective function sought to be minimized in CH-partitioning is the squared sum of core and halo nodes. Since in the above definition, $h_{vi}=1$ if $v$ is either a core or a halo vertex, it suffices to minimize $\left( \sum_{v \in V} h_{vi} \right)^2$, the squared sum of halos for each partition $i \in \{1,\ldots,K\}$, expressed in $H_C$.

The weights $A$, $B$ and $C$ have to be chosen appropriately again to ensure the correct global minimum of $H$ at a valid CH-partitioning. This is done as follows. Suppose $H$ is in equilibrium at a given configuration of $c_{vi}$, $h_{vi}$ and $z_{(v,w),i}$. First, assume we change a $c_{vi}$ from $0$ to $1$. This will not happen in practice since $c_{vi}=1$ potentially only results in more $h_{wi}=1$ (for $w \in N(v) \cup \{ v\}$), thus increasing the objective function $H_C$. Assume now that $c_{vi}$ changes from $1$ to $0$. Since $H_A=0$ in equilibrium, changing one $c_{vi}$ results in $\Delta H_A \geq 1$. Although the objective function $H_C$ does not depend on any $c_{vi}$, changing a $c_{vi}$ can still lead to a decrease of the objective. This is due to the fact that switching $c_{vi}$ from $1$ to $0$ potentially causes $\Delta$ (the maximal degree of the graph) halo indicators to not be forced to be one any more, hence leading to a maximal decrease of $H_C$ of $\Delta H_C \leq \left( \sum_{v \in V} h_{vi} \right)^2 - \left( \sum_{v \in V} h_{vi} - \Delta \right)^2 \leq |V|^2$. Together, we see that $A \Delta H_A \geq A > C |V|^2 \geq C \Delta H_C$, hence $A > C|V|^2$. Setting $C=1$ leads to the choice $A=|V|^2+1$.

Likewise, consider changing a halo indicator $h_{vi}$ out of equilibrium. Assume $h_{vi}$ changes from $0$ to $1$. In this case, we would increase the penalty $H_B$ as well as the objective function $H_C$, meaning this case is infeasible in practice. Assume now that we change $h_{vi}$ from $1$ to $0$. In this case, we incur an increase in $H$ resulting from the penalty term $H_B$ of at least $\Delta H_B \geq 1$ (since we can assume that in equilibrium, $c_{vi}=1$ whenever $h_{vi}=1$). The term $H_A$ is independent of $h_{vi}$ and hence not affected. The objective function $H_C$ will decrease by at most $\Delta H_C \leq |V|^2 - (|V|-1)^2$ in the worst case if one $h_{vi}$ switches from $1$ to $0$, hence $\Delta H_C \leq 2|V|-1 \leq 2|V|$. Together, we see that $\Delta H_B \geq B > 2|V|C \geq \Delta H_C$, thus $B > 2|V|C$. Using $C=1$ set before results in the choice $B = 2|V|+1$.

\subsection{Simulation results}
\label{section_simulations}
The following section shows simulation results for an evaluation of classical edge-cut graph partitioning on D-Wave (both using the Hamiltonian for two partitions given in Section \ref{section_hamiltonians_ising} as well as the one of Section \ref{section_hamiltonians_qubo}) and for CH-partitioning (Section \ref{section_hamiltonians_ch}). We compare D-Wave to \texttt{METIS} \citep{KarypisKumar1999}, to simulated annealing as well as to \texttt{Gurobi} (see Section \ref{section_classical_solvers} for details and/or the precise algorithm). All D-Wave solvers described in Section \ref{section_dwave_solvers} are included in our study.

We start by describing our simulation setting and proceed to evaluate the D-Wave performance on edge-cut partitioning as well as CH-partitioning. Since graph which can be embedded on the D-Wave chip are rather small, we switch to graph partitioning into two components for maximal sized graphs. The section finishes with an evaluation of the embedding vs.\ computing time on D-Wave.

\subsubsection{Setting and choice of test systems}
\label{section_simulations_testsystems}
All graph partitioning methods were evaluated on simulated graphs consisting of $6$ to $48$ vertices. For all simulated graphs, we inserted random edges between all pairs of vertices with probability $0.9$. These graphs remained fixed throughout the comparison.

The largest graph size we were able to use was limited by the size of the corresponding Hamiltonian that could still be embedded on the D-Wave chip. However, solely the Sapi method for D-Wave (see Section \ref{section_dwave_solvers}) has this constraint -- the two other D-Wave solvers (QBSolv and QSage) are (classical) hybrid methods which can handle Hamiltonians of arbitrary size.

Unless otherwise stated, all evaluations presented hereafter compare the classical methods \texttt{METIS}, simulated annealing (SA) and \texttt{Gurobi} to the D-Wave solvers Sapi, QBSolv and QSage. QBSolv and QSage are straightforward to use since they do not allow for significant parameter tuning. For Sapi, we re-compute the graph embedding onto the D-Wave chip in each run in order to maximize the Hamiltonian size solvable on D-Wave, always perform $1000$ anneals, and use the \textit{answer\_mode='raw'} as well as the \textit{embeddings='chains'} and \textit{broken\_chains='minimize\_energy'} post processing modes.

Computing times were measured as averages of $1000$ repetitions. For D-Wave, we reported the \textit{total\_real\_time} provided by D-Wave's \textit{solve\_ising} method. This time does not include the time needed to compute embeddings of the Hamiltonian onto D-Wave's chimera chip.

\subsubsection{Edge-cut partitioning}
\label{section_simulations_ec}
\begin{figure}
\centering
\includegraphics[width=0.5\textwidth]{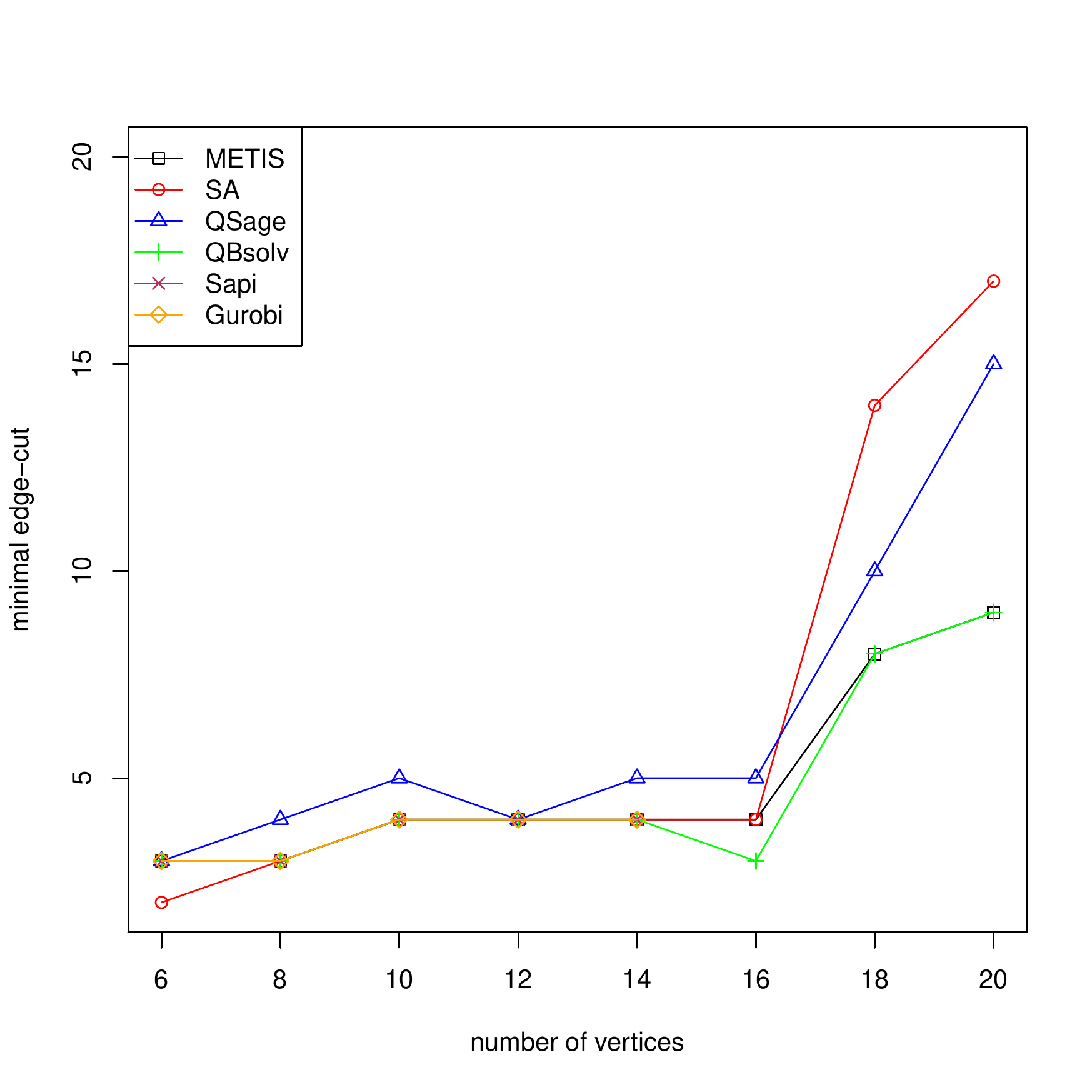}
\caption{Edge-cut partitioning (in four components): Minimal edge-cut found by the three classical and three D-Wave methods employed in this study as a function of the number of vertices in the test graph.\label{fig:georg_ec_ec}}
\end{figure}
\begin{figure}
\centering
\includegraphics[width=0.5\textwidth]{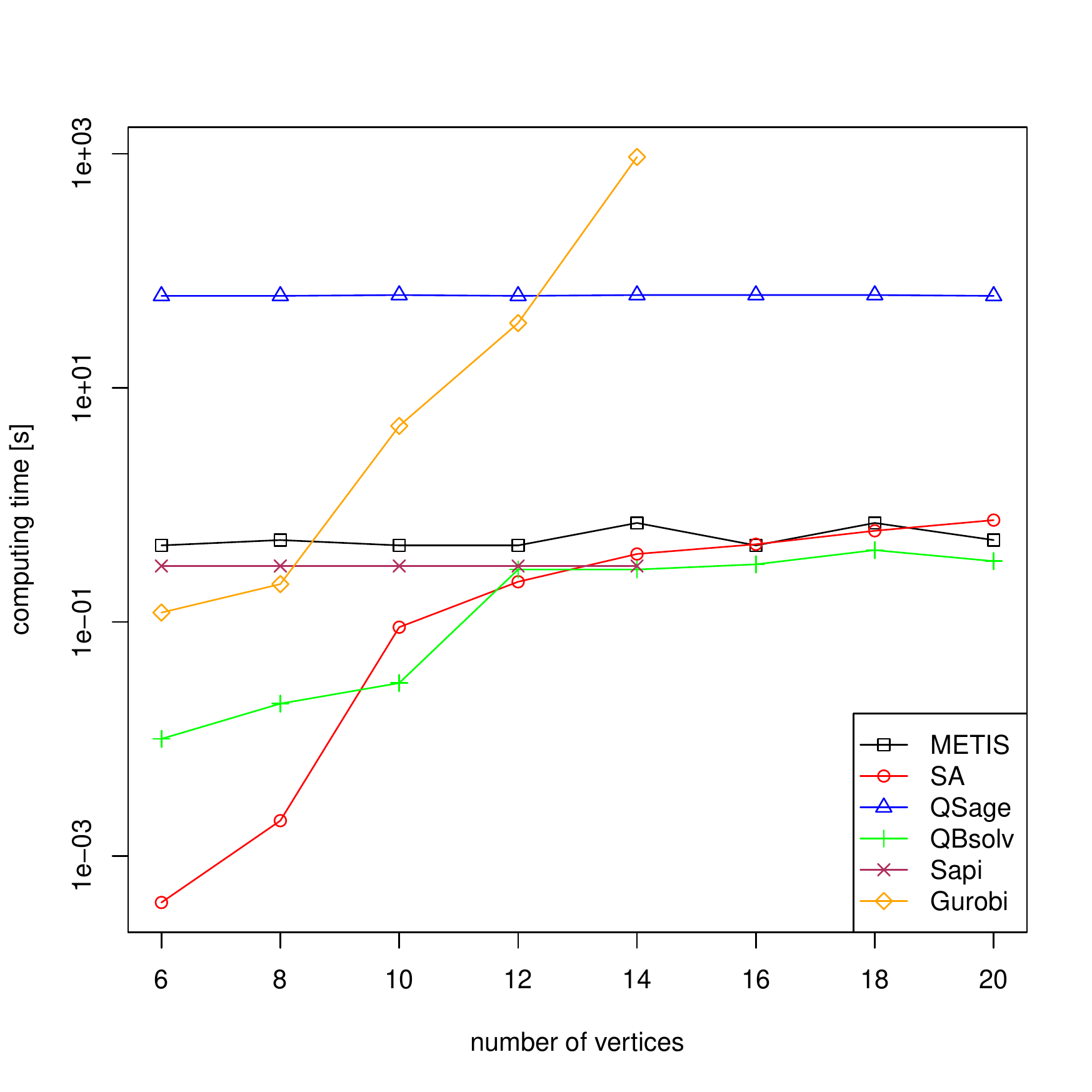}
\caption{Edge-cut partitioning (in four components): Average computing time for all six methods to find a solution of the quality shown in Figure \ref{fig:georg_ec_ec}.\label{fig:georg_ec_time}}
\end{figure}
We start by evaluating D-Wave's performance on classical edge-cut graph partitioning. For added difficulty we aim to split graphs into four balanced partitions. Since we aim to split graphs into more than two partitions we use the Qubo Hamiltonian derived in Section \ref{section_hamiltonians_qubo} and apply classical and D-Wave solvers to the set of graphs fixed in Section \ref{section_simulations_testsystems}.

Figure \ref{fig:georg_ec_ec} shows the best solution found among $1000$ repetitions (for the classical methods) and $1000$ anneals for Sapi and QSage. QBSolv generally produces very stable results, meaning that repeating computations with QBSolv generally does not lead to varying answers.

As shown in Figure \ref{fig:georg_ec_ec}, all methods perform comparably well on the test systems, especially for small graphs. For larger graphs (towards $20$ vertices), \texttt{METIS} and QBSolv find the best partitionings (in terms of a minimal edge-cut), SA performs well for small graphs but increasingly worse for larger graphs. However, the implementation we use for SA is a generic simulated annealing algorithm not further optimized (for instance, by choosing specific proposals or a certain temperature function). QSage seems to perform slightly worse than the other methods throughout the range of test graphs.

Figure \ref{fig:georg_ec_time} shows the corresponding average computing time to Figure \ref{fig:georg_ec_ec}. As can be seen from Figure \ref{fig:georg_ec_time}, SA is the fastest method (on average), thus providing very good results at short times for small graphs, and less good results for larger graphs.

As QSage is run with a time-out criterion (set to $60$ seconds in our experiments), and as it is not possible to determine for QSage when a global minimum has been reached, using up its full computing time in each run. The time-out of $60$ seconds is an arbitrary choice (although for too short timethe s, results will start getting worse).

Notably, \texttt{METIS} is able to compute very good solutions in around $0.4$ seconds throughout the comparison.

The time for Sapi is constant at $0.3$ seconds. This is the overall time for computation and post-processing on the D-Wave server side for $1000$ anneals from D-Wave. However, Sapi is unable to embed graph Hamiltonians of the form derived in Section \ref{section_hamiltonians_qubo} beyond $14$ vertices.

\texttt{Gurobi} allows to compute partitionings which are proven to be globally optimal. However, its running time is exponential, thus greatly limiting its range of applicability. We were not able to compute partitionings with \texttt{Gurobi} beyond $14$ vertices.

Overall, we conclude from these results that Sapi/D-Wave does not yet offer a quantum advantage over classical methods since the results obtained with QBSolv make heavy use of classical post-processing algorithms.

\subsubsection{CH-partitioning}
\label{section_simulations_ch}
\begin{figure}
\centering
\includegraphics[width=0.5\textwidth]{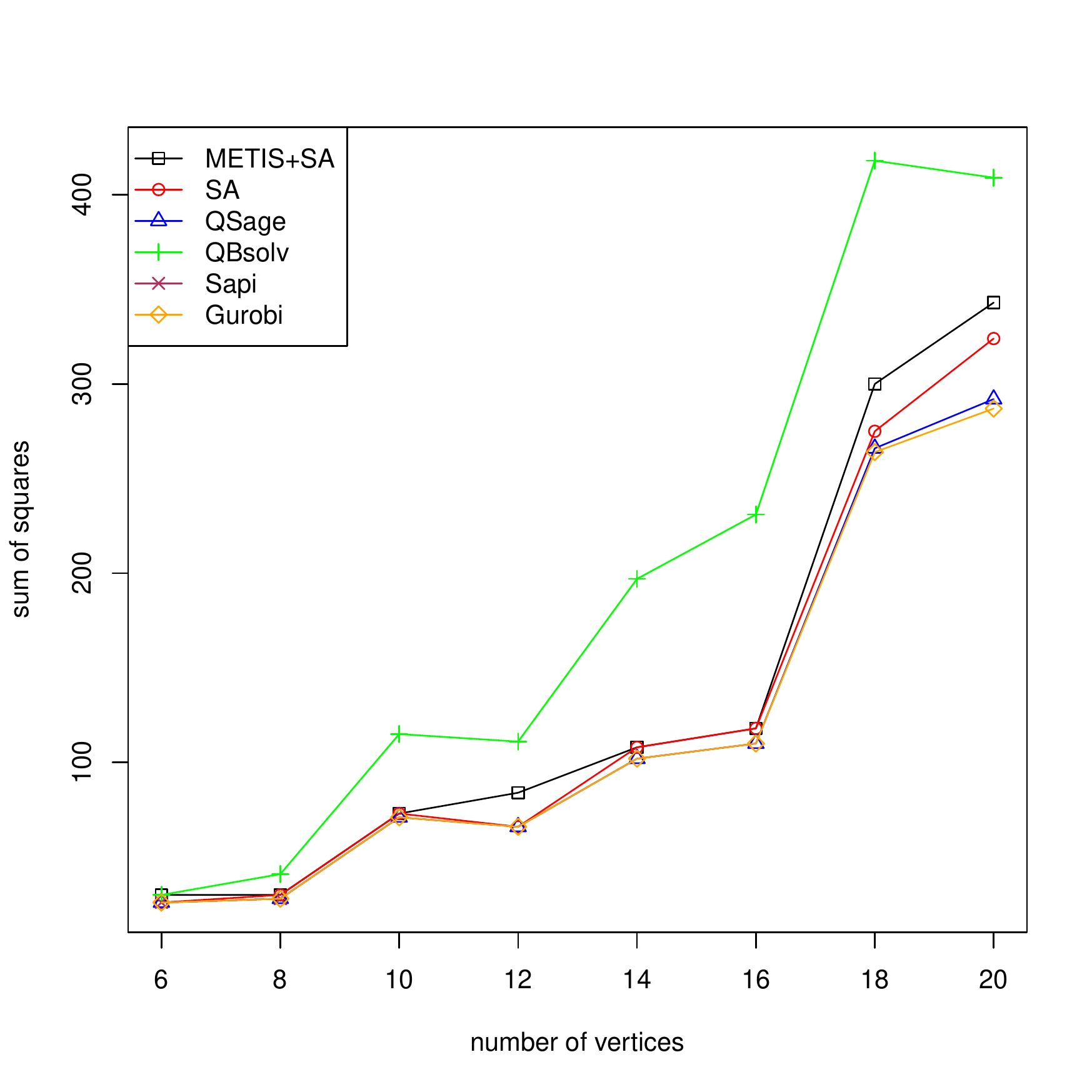}
\caption{CH-partitioning: Partition allocation minimizing the sum of squares criterion found by the three classical and three D-Wave methods employed in this study as a function of the number of vertices in the test graph.\label{fig:georg_sumsq_sumsq}}
\end{figure}
\begin{figure}
\centering
\includegraphics[width=0.5\textwidth]{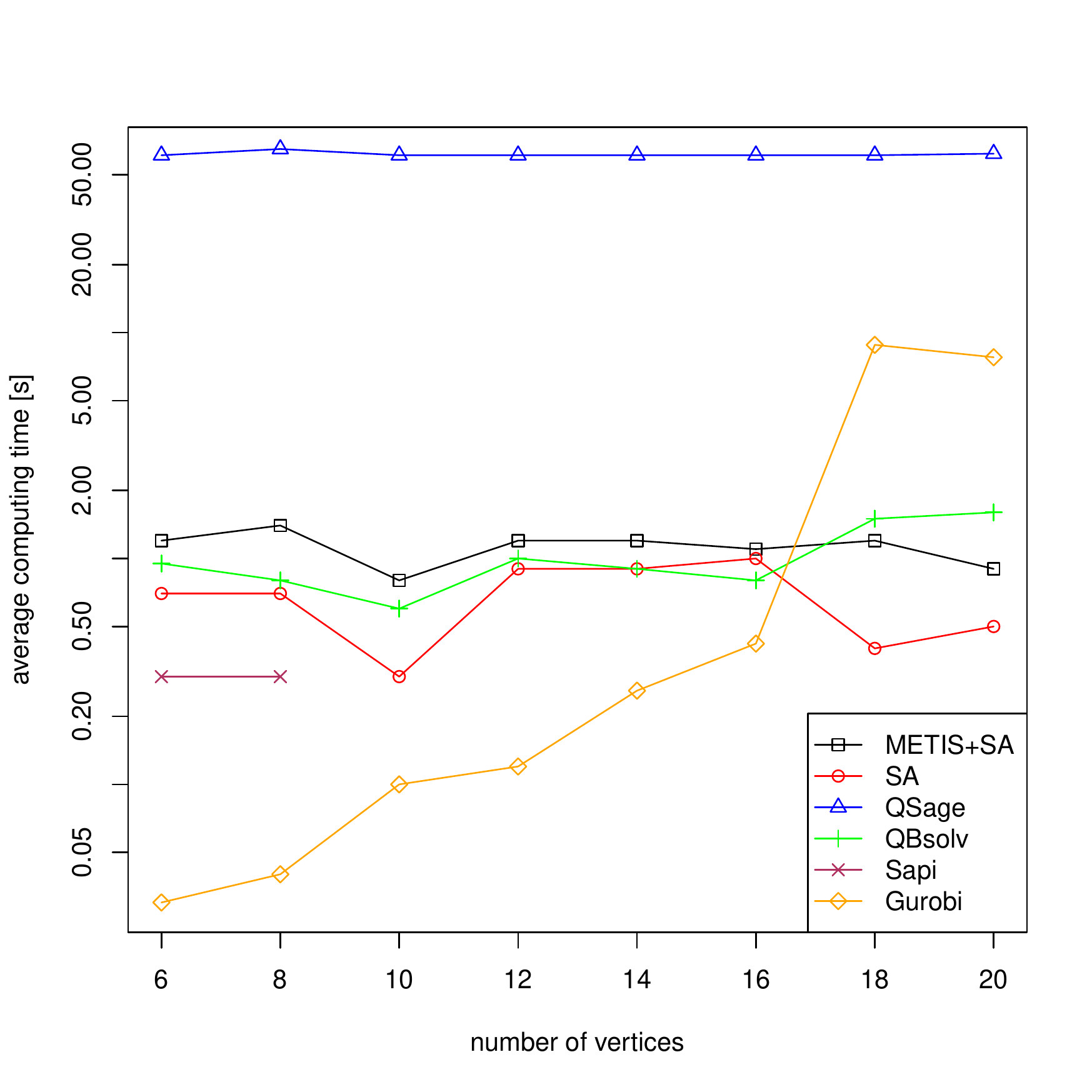}
\caption{CH-partitioning: Average computing time for all six methods to find a solution of the quality shown in Figure \ref{fig:georg_sumsq_sumsq}.\label{fig:georg_sumsq_time}}
\end{figure}
Next, we repeat the previous comparison for CH-partitioning (again into four partitions) using the same test graphs and the Hamiltonian derived in Section \ref{section_hamiltonians_qubo}. The method abbreviated by SA in this section is a simulated annealing algorithm tailored to CH-partitioning proposed in \cite{Djidjev2016} (see Section \ref{section_classical_solvers_METIS_SA}). We employ this SA algorithm both as a post-processing step to tune partitions computed by \texttt{METIS} (which minimize the edge-cut instead of \eqref{eq:sum_of_squares}) and as a stand-alone algorithm initialized with random partitions.

Figure \ref{fig:georg_sumsq_sumsq} shows simulation results. Precisely, we evaluate the partitions found for each method and graph size by plotting the sum of squares criterion \eqref{eq:sum_of_squares} (Section \ref{section_hamiltonians_ch}), where $c_i$ is the partition (core) size for each partition $i \in \{ 1,\ldots,K\}$, and $h_i$ is the number of neighbors in adjacent partitions (the halo). As visible from the plot, performance is more similar for all methods under investigation apart from QBSolv, which seems to perform poorly.

Notably, \texttt{METIS} with a SA post-processing step performs worse than using SA on its own with a random initial partitioning allocation. QSage finds very good CH-partitions, albeit having a very long $60$ second running time (Figure \ref{fig:georg_sumsq_time}). As before, \texttt{Gurobi} finds optimal partitions in the sense that they globally minimize the sum of squares criterion, even though problems become infeasible to solve exactly as the number of vertices increases. Unfortunately, Sapi is only able to embed the CH-partitioning Hamiltonian for $6$ and $8$ vertex graphs. This is due to the fact that CH-partitioning uses a large number of auxiliary variables for each partition and vertex (see Section \ref{section_hamiltonians_ch}), thus resulting in relatively large Hamiltonians even for small graph sizes.

Figure \ref{fig:georg_sumsq_time} shows corresponding computing times. The three classical methods (\texttt{METIS} and SA, SA by itself and QBSolv) perform comparably with a decent running time of around one second. Due to the user-specified time-out criterion of $60$ seconds, QSage again has a constant running time. Sapi again achieves a fast $0.3$ second anneal time but is, however, only applicable to small graph instances. \texttt{Gurobi} achieves the fastest runtimes for small instances among all methods considered, nevertheless its runtime quickly falls behind the other methods (apart from Sapi).

\subsubsection{Edge-cut partitioning with Ising Hamiltonian}
\label{section_simulations_ising}
\begin{figure}
\centering
\includegraphics[width=0.5\textwidth]{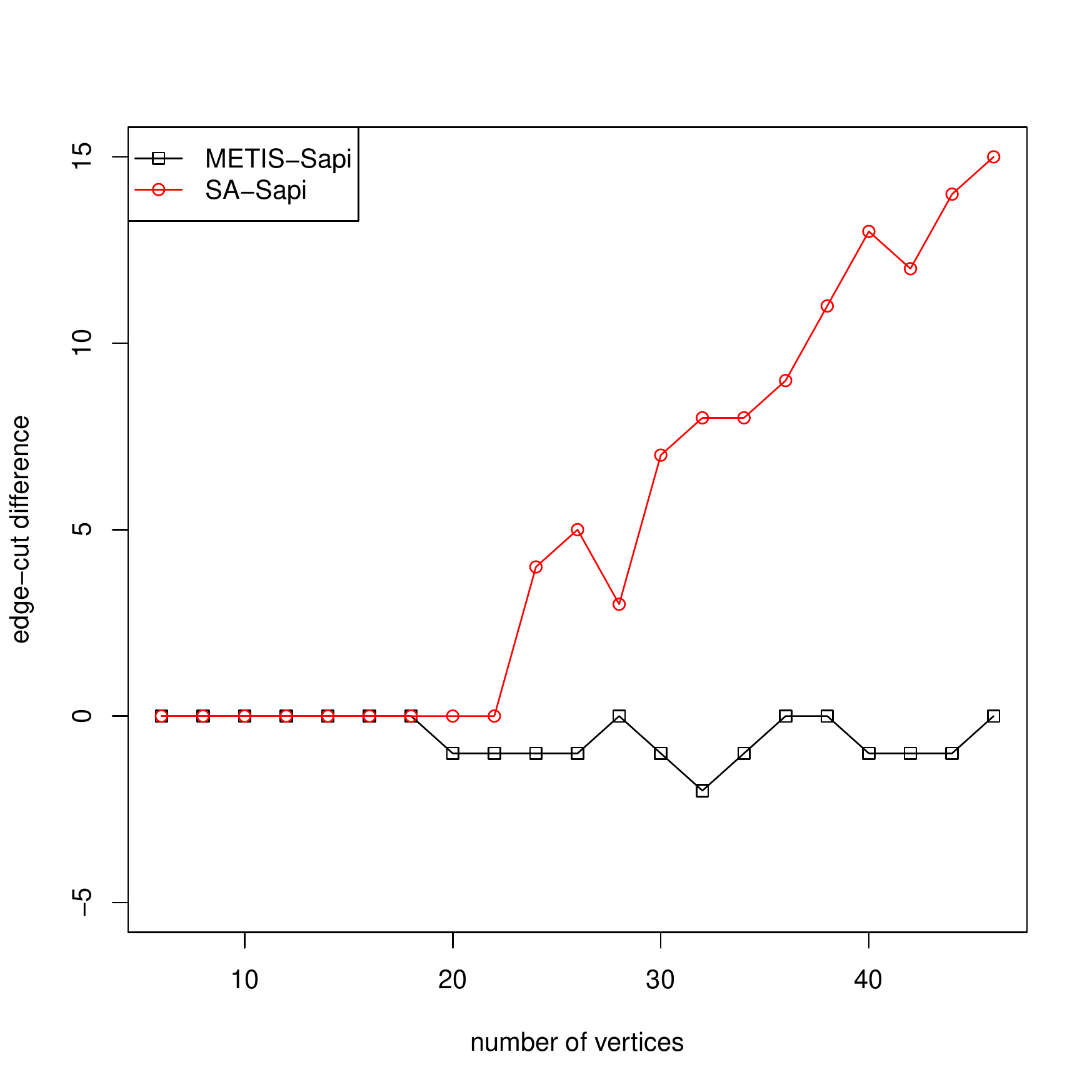}
\caption{Edge-cut partitioning with Ising Hamiltonian (in two components): Differences in the minimal edge-cut found by D-Wave's Sapi and \texttt{METIS} or simulated annealing (SA) as a function of the number of vertices in the test graph.\label{fig:georg_2part_ec}}
\end{figure}
\begin{figure}
\centering
\includegraphics[width=0.5\textwidth]{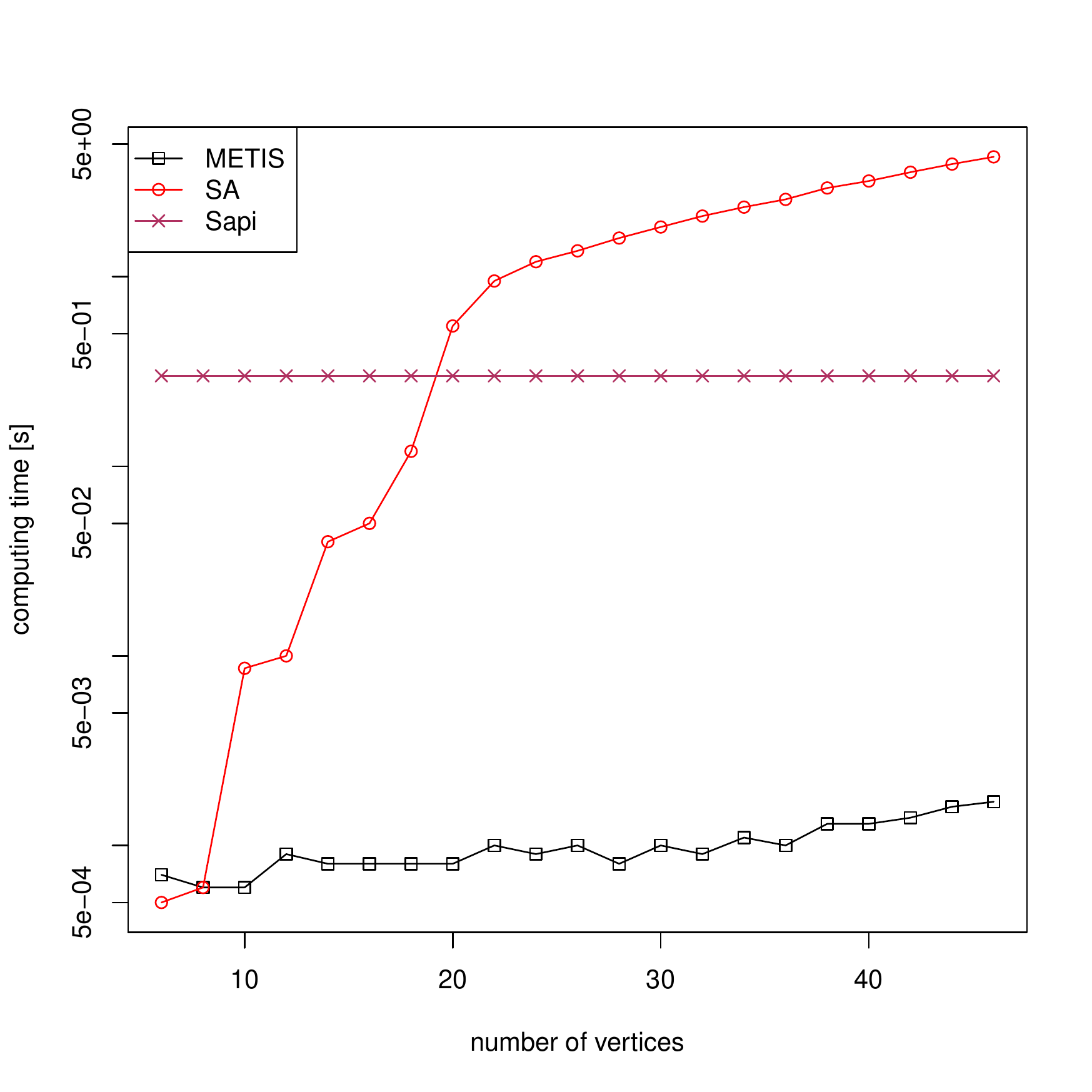}
\caption{Edge-cut partitioning with Ising Hamiltonian (in two components): Computing time for D-Wave's Sapi, \texttt{METIS} and simulated annealing (SA) corresponding to Figure \ref{fig:georg_2part_ec} as a function of the number of vertices in the test graph.\label{fig:georg_2part_time}}
\end{figure}
Section \ref{section_simulations_ec} already investigated edge-cut partitioning with four components. It established that D-Wave's QBSolv performs as well as classical methods, though with the help of classical post-processing to enhance the initial solution returned by the D-Wave device. Unfortunately, the tool which provides closest quantum control, Sapi, was not able to embed graph instances of sufficient size.

This is due to two factors: First, splitting a graph into more partitions uses up more qubits to indicate which partition each vertex belongs to. Thus, evaluating graph partitioning with two partitions allows for a maximal number of vertices in the input graph. Second, the Hamiltonian for graph partitioning into an arbitrary number of partitions derived in Section \ref{section_hamiltonians_qubo} uses as many qubits per vertex as there are partitions, meaning that even for two partitions, two qubits per vertex are used. This unnecessarily uses up qubits and hence limits the size of the input graph. For the special case of two partitions, the Ising formulation of Section \ref{section_hamiltonians_ising} allows to allocate one qubit only per vertex.

Since for small instances, classical (brute force) methods are at least as fast as sophisticated (quantum or classical) methods, we re-evaluate graph partitioning with Ising Hamiltonian to allow for a more pronounced quantum speedup.

Figure \ref{fig:georg_2part_ec} shows simulation results for graphs up to $48$ vertices (qubits in Ising formulation), the largest graph size which can be embedded on the D-Wave chip. We depict the difference in the best edge-cut found by \texttt{METIS}, simulated annealing (as classical solvers) in relation to D-Wave's Sapi, since Sapi offers closest control over D-Wave. The SA algorithm we employ in this section is the one given as Algorithm \ref{algorithm_sa} in Section \ref{section_classical_solvers_SA}. The figure shows that all methods are able to find the same minimal edge-cut for instances up to $18$ vertices. For larger graphs, however, D-Wave's Sapi considerably outperforms classical simulated annealing and finds partitions which even improve upon the ones found by \texttt{METIS} by up to edge-cut two.

The corresponding computing times for all three methods are displayed in Figure \ref{fig:georg_2part_time}. As before, it shows that D-Wave/ Sapi has a constant computing time of around $0.3$ seconds, while \texttt{METIS} produces good solutions in a very short time. SA runs very fast for small graphs (up to $18$ vertices) but performs considerably poorer in terms of computing time for larger instances.

\subsubsection{Timings for embedding and anneal steps}
\label{section_simulations_timings}
\begin{figure}
\centering
\includegraphics[width=0.5\textwidth]{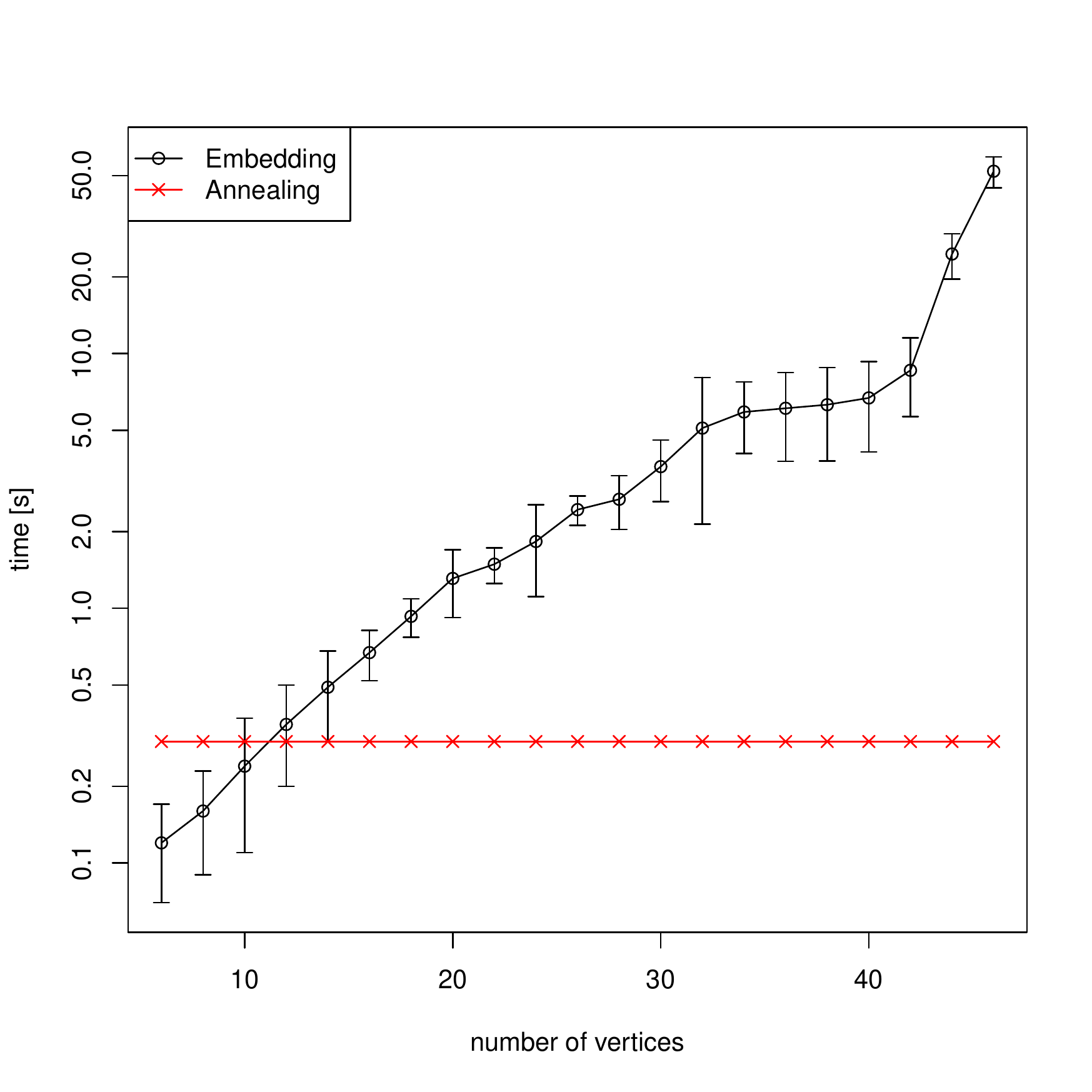}
\caption{Embedding time (black) and anneal time (red) for the fixed set of test graphs up to $48$ vertices using the Ising Hamiltonian of Section \ref{section_hamiltonians_ising}. Circles indicate means computed with $10$ repetitions.\label{fig:georg_emb_time}}
\end{figure}
In order to solve a Hamiltonian on the D-Wave chip, it is necessary to embed it first on its \textit{chimera} architecture. This has to be done once for each Hamiltonian and can be quite time consuming.

Using the set-up of Section \ref{section_simulations_ising} (graph partitioning with Ising Hamiltonian into two partitions using our set of test graphs), we compare the timings to compute graph embeddings to the (constant) anneal times for $1000$ samples from D-Wave. Figure \ref{fig:georg_emb_time} reports mean timings (circles) for $10$ repetitions together with error bars indicating the variance. It shows that the time to compute the embeddings roughly increases linearly with the graph (Hamiltonian) size. For small instances, the embedding time is negligible. However, for large instances the embedding time quickly exceeds the actual anneal (computing) time on D-Wave, thus limiting the capabilities of D-Wave.

Alternatively, a straightforward way to mitigate this problem would be to always use pre-computed embeddings for full graphs: D-Wave allows to use pre-compute embeddings for complete graphs of size up to $45$ vertices, hence allowing to use these embeddings for all graphs up to that vertex size. Nevertheless, computing individual embeddings allows to use qubits which are physically close on the D-Wave chip, thus minimizing chain lengths which increases computation accuracy.

\subsubsection{Histogram of maximal clique sizes}
\label{section_simulations_histogram_cliques}
\begin{figure}
\centering
\includegraphics[width=0.5\textwidth]{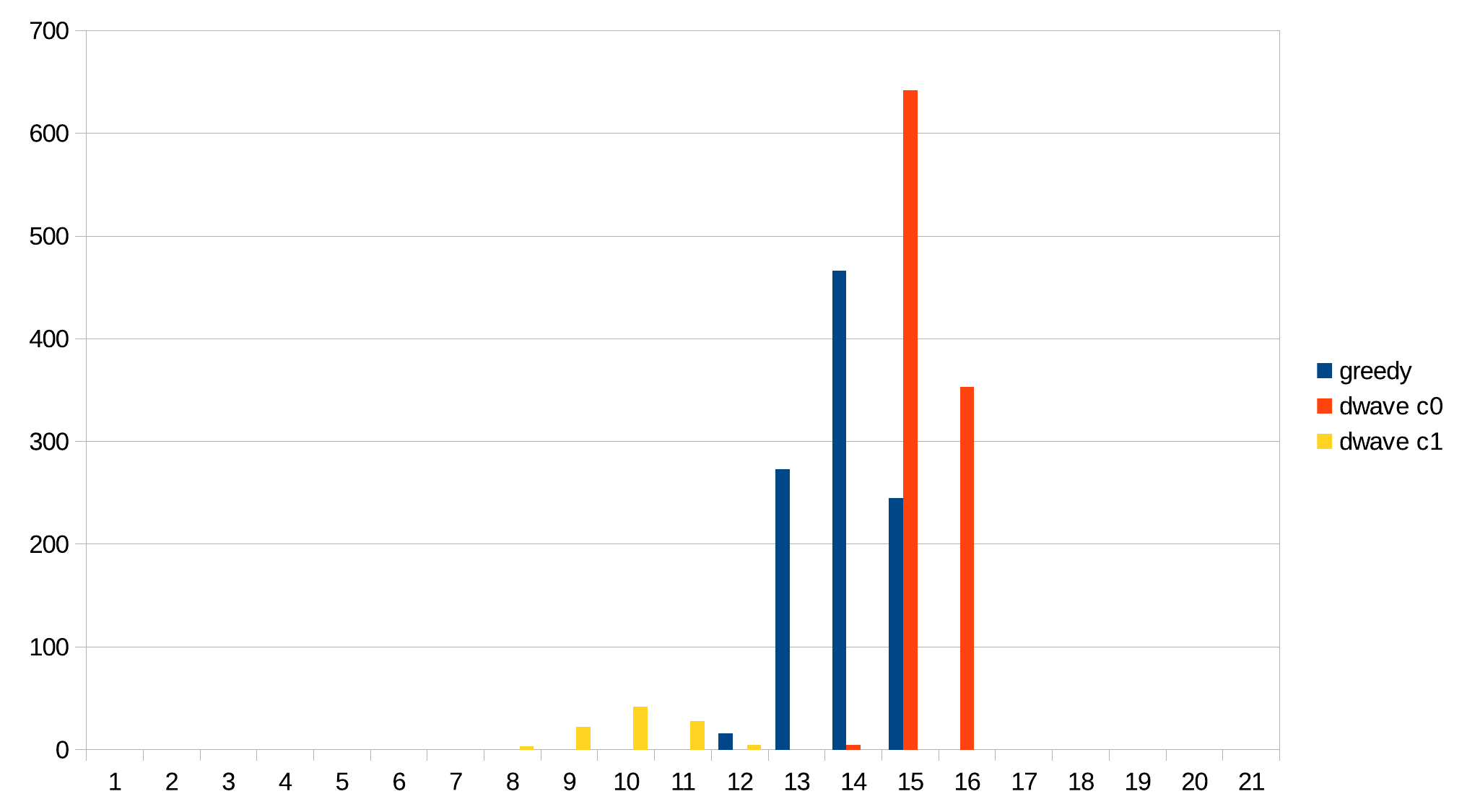}
\caption{Histogram showing the maximal clique sizes found by D-Wave and a greedy algorithm. D-Wave is used with two options, once with chain strength $0$ and once with chain strength $-1$.\label{fig:georg_hist}}
\end{figure}
Figure~\ref{fig:georg_hist} shows a histogram of maximal clique sizes found by D-Wave and a greedy algorithm. D-Wave is used with two options, once with chain strength $0$ and once with chain strength $-1$.

\section{Conclusions}
\label{section_conclusion}
This article evaluated the performance of the D-Wave 2X quantum annealer on two NP-hard graph problems, clique finding and graph partitioning.

We compared D-Wave's solvers QSage, QBSolv and Sapi to common classical solvers (quantum adiabatic evolution of \cite{childs2000finding}, specialized simulated annealing of \cite{geng2007simple}), \texttt{fmc} of \cite{pattabiraman2013fast}, and \texttt{Gurobi} for clique finding; \texttt{METIS} \citep{KarypisKumar1999}, two simulated annealing approaches of \cite{KadowakiNishimori1998,Djidjev2016} and \texttt{Gurobi} for graph partitioning) to determine if current technology already allows to observe a \textit{quantum advantage} in these areas. We summarize our findings as follows.
\begin{enumerate}
  \item Problems need to be of a certain (problem-dependant) size to see a performance gain with D-Wave. General problems can often not be embedded on the D-Wave chip; if they can be embedded, however, D-Wave returns solutions of comparable quality than classical ones, even though classical methods are usually faster for such small instances.
  \item For random larger instances which were designed to fit the D-Wave architecture, a speed-up for D-Wave in comparison to classical solvers is observable. However, it is not straightforward to justify the correct time to actually measure on D-Wave.
  \item The graph embedding needed to map a particular Hamiltonian to the qubits on the D-Wave chip is computationally expensive and hence time-consuming, thus considerably distorting the time measurements if added to the actual anneal (solver) time.
  \item In comparison to our simulated annealing (SA) implementations, D-Wave perfroms considerably better. However, it is not clear if such a comparison is fair since our implementation of SA might not be sufficiently optimized and might be outperformed by any hardware solver, either classical or quantum.
\end{enumerate}
Overall, we conclude that general problems which allow to be mapped onto the D-Wave architecture are typically still too small to show a quantum advantage. Such instances can also often be solved faster (and often even exact) with best available classical algorithms (returning solutions of at least the quality that D-Wave returns). In comparison to simple SA algorithms, D-Wave is considerably faster, even though we claim no optimization for our SA routines. Selected instances especially designed to fit D-Wave's particular \textit{chimera} architecture can be solved orders of magnitude faster with a quantum annealer than with classical techniques.


\end{document}